\documentclass[
	aps, prd,  notitlepage, 
        floats, floatfix, twocolumn,
	amsmath, amssymb, amsfonts, eqsecnum,
	superscriptaddress,
	showpacs, showkeys,
	nofootinbib,
 	longbibliography,
]{revtex4-1}

\usepackage{graphicx}
\usepackage[usenames,dvipsnames]{xcolor}
\usepackage{tikz}
\usepackage{xspace} 
\usepackage{bm}
\usepackage[utf8]{inputenc} 
\usepackage{latexsym}
\usepackage{float}
\usepackage{ulem}
\renewcommand{\emph}[1]{\textit{#1}}
\xdefinecolor{mylinkcolor}{rgb}{0,0,0.5}
\usepackage[
	bookmarksnumbered, bookmarksopen, bookmarksopenlevel=2,
	breaklinks=true,
	colorlinks=true, filecolor=mylinkcolor, citecolor=mylinkcolor,
	linkcolor=mylinkcolor, urlcolor=mylinkcolor, menucolor=mylinkcolor,
]{hyperref}

\def\be{\begin{equation}}
\def\ee{\end{equation}}
\def\bea{\begin{eqnarray}}
\def\eea{\end{eqnarray}}
\newcommand{\bes}{\begin{subequations}}
\newcommand{\ees}{\end{subequations}}

\def\di{\partial}

\allowdisplaybreaks

\begin{document}

\title{Foundations of an effective-one-body model for coalescing binaries on eccentric orbits}
\date{\today}

\author{Tanja Hinderer}
\affiliation{Max Planck Institute for Gravitational Physics (Albert Einstein Institute), Am M\"uhlenberg 1, Potsdam, 14476, Germany}
\affiliation{Department of Astrophysics/IMAPP, Radboud University, P.O. Box 9010,
6500 GL Nijmegen, The Netherlands}

\author{Stanislav Babak}
\affiliation{Max Planck Institute for Gravitational Physics (Albert Einstein Institute), Am M\"uhlenberg 1, Potsdam, 14476, Germany}
\affiliation{APC, Univ. Paris Diderot, CNRC/IN2P3, CEA/lrfu, Obs. de Paris, Sorbonne Paris Cite, France}
\affiliation{Moscow Institute of Physics and Technology, Dolgoprudny, Moscow region, Russia}

\begin{abstract}
We develop the foundations of an effective-one-body (EOB) model for eccentric binary coalescences that includes the conservative dynamics, radiation reaction, and gravitational waveform modes from the inspiral and the merger-ringdown signals. Our approach uses the strategy that is commonly employed in black-hole perturbation theory: we introduce an efficient, relativistic parameterization of the dynamics that is defined by the orbital geometry and consists of a set of phase variables and quantities that evolve only due to gravitational radiation reaction. Specializing to nonspinning binaries, we derive the EOB equations of motion for the new variables and make use of the fundamental frequencies of the motion to compute the binary's radiative multipole moments that determine the gravitational waves. Our treatment has several advantages over the quasi-Keplerian approach that is often used in post-Newtonian (PN) calculations: a smaller set of variables, parameters that reflect the features of strong-field dynamics, and a greater transparency of the calculations when using the fundamental frequencies that leads to simplifications and an unambiguous orbit-averaging operation. While our description of the conservative dynamics is fully relativistic, we limit explicit derivations in the radiative sector to $1.5$PN order for simplicity. This already enables us to establish methods for computing both instantaneous and hereditary contributions to the gravitational radiation in EOB coordinates that have straightforward extensions to higher PN order. The weak-field, small eccentricity limit of our results for the orbit-averaged fluxes agrees with known PN results when expressed in terms of gauge-invariant quantities. We further address considerations for the numerical implementation of the model and the completion of the waveforms to include the merger and ringdown signals, and provide illustrative results. 
\end{abstract}

\maketitle

\section{Introduction}
The recent first detections of gravitational waves (GWs) from merging black holes (BHs) by the advanced Laser Interferometer Gravitational-wave Observatory (LIGO) have initiated the vibrant field of GW astronomy~\cite{Abbott:2016blz,TheLIGOScientific:2016pea}. A large sample of binary BH observations is anticipated to accumulate as LIGO's sensitivity improves and a worldwide network of detectors (Virgo~\cite{TheVirgo:2014hva}, KAGRA~\cite{Aso:2013eba}, and LIGO India \cite{LIGOIndia}) becomes operational. These observations will enable unprecedented tests of general relativity and the nature of BHs. Furthermore, they will provide invaluable astrophysical information on the endpoints of the evolution of massive stars and the formation channels of compact-object binary systems, their evolution, and astrophysical environments. Binary BHs are also major targets for the planned space-based detector Laser Interferometer Space Antenna (LISA)~\cite{lisa}. LISA is expected to observe merging supermassive BHs throughout cosmic time, at all important epochs in their evolutionary history~\cite{Klein:2015hvg}, as well as small mass ratio inspirals from within the deepest regions of galactic nuclei and the strong-field spacetime of supermassive BHs~\cite{Babak:2017tow}.

The unique information encoded in the GW signals from binaries is extracted by using matched filtering. This method cross-correlates the detector output with a bank of theoretical predictions for GW signals (templates) for a wide range of possible parameters characterizing the binary system. Accurate template models that include all relevant physical effects in a merging binary system are therefore essential to detect weak signals and maximize the science payoffs from the observations. 

Current state-of-the art template models for ground-based detectors describe binaries on quasi-circular orbits~\cite{Bohe:2016gbl,Khan:2015jqa,Pan:2013rra,Hannam:2013oca}. Circular orbits are expected for binaries that formed at large separation from a progenitor binary stellar system, where longterm GW losses have rapidly reduced the eccentricity~\cite{PhysRev.131.435}. However, there exist several mechanisms through which BH binaries may retain a significant eccentricity when entering the sensitive frequency band of ground-based detectors. In dense stellar environments, such as globular clusters and galactic centers, dynamical friction causes BHs to segregate towards the core of the cluster or galaxy, where frequent dynamical interactions with stars and other BHs can lead to the formation of eccentric BH binaries at small orbital separation. This can occur either through direct dynamical capture or in hierarchical triple systems, where the Kozai-Lidov mechanism \cite{1962AJ.....67..591K, 1962P&SS....9..719L} can secularly drive the inner binary to high eccentricity. The merger rate for these binaries is uncertain but could be a few events per year for advanced LIGO~\cite{Antonini:2015zsa, Samsing:2016bqm,VanLandingham:2016ccd}. A further possible mechanism to induce eccentricity is through  the kick from the supernova explosion forming the secondary compact object that, depending on the direction and magnitude, may lead to a residual eccentricity at small separation. Additional potential sources for terrestrial GW observatories where eccentricity could play a role are the inspirals of stellar-mass BHs into intermediate-mass BHs~\cite{Brown:2006pj}. Although eccentric inspirals may be rare events for the advanced GW detector network, their observation will reveal a highly interesting population of binaries whose GW signals will be richer in structure than those from binaries on circular orbits. Conversely, non-detections of eccentric systems will significantly constrain previously inaccessible astrophysics. To realize these science benefits requires accurate template models that include features due to eccentricity. Moreover, models with eccentricity are needed to robustly determine that orbits are quasi-circular and thereby reduce systematic errors in all measured parameters, which is an important prerequisite for probing fundamental physics. 
The main importance of template models that include features for arbitrary eccentricity is for observations with LISA. Both classes of LISA's most interesting sources, the merging supermassive BHs~\cite{Klein:2015hvg} and the small mass ratio inspirals~\cite{Babak:2017tow}, will generically have eccentric orbits and require appropriate models.

There have been several previous studies on computing GWs from eccentric binaries. For extreme mass ratio systems, snapshot adiabatic waveforms have been computed from black-hole perturbation theory both numerically~\cite{Drasco:2005kz}, from fits to the numerical fluxes that enable efficient computations of approximate inspirals~\cite{Babak:2006uv}, and from series solutions to the black-hole perturbation equations~\cite{Forseth:2015oua}. Approximate inspirals that include the full first order gravitational self-force, describing the small mass' interaction with its own spacetime distortion, have also been computed~\cite{Warburton:2011fk,Osburn:2015duj}. A more comprehensive list of references on related work on extreme mass ratio inspirals can be found in the review articles in Refs.~\cite{Babak:2017tow, 2009CQGra..26u3001B,2015JPhCS.610a2002A,2011LRR....14....7P}. In the context of comparable-mass binaries, a number of studies have focused on computing the dynamics and gravitational radiation in post-Newtonian (PN) theory; see 
Refs.~\cite{Memmesheimer:2004cv, Arun:2007rg, Mishra:2015bqa, Arun:2009mc, Arun:2007sg, Boetzel:2017zza} for the most recent updates and other references, Refs.~\cite{Loutrel:2016cdw,Loutrel:2017fgu} for explicit approximations to the PN fluxes for arbitrarily high eccentricity, and 
Refs.~\cite{Tanay:2016zog,Damour:2004bz,Konigsdorffer:2006zt,Moore:2016qxz,2009PhRvD..80h4001Y,huerta2014accurate} for the GW phasing in the limit of small eccentricity. Comparisons have also been performed between the small mass ratio and the PN approximation~\cite{Akcay:2015pza}, and with numerical-relativity studies~\cite{Lewis:2016lgx}. Complete waveform models for data analysis that go beyond the PN description of the inspiral have recently also been developed~\cite{Huerta:2016rwp,Hinder:2017sxy}. These models describe binaries with a small orbital eccentricity either by connecting a PN inspiral augmented with additional knowledge from gravitational self-force results to a phenomenological merger-ringdown model~\cite{Huerta:2016rwp}, or by matching PN inspiral waveforms onto circular-orbit numerical relativity (NR) mergers~\cite{Hinder:2017sxy}.

The purpose of this paper is to develop theoretical and practical foundations to describe generic inspirals in the Effective-One-Body (EOB) model~\cite{Buonanno:1998gg}. The EOB approach provides a framework to combine PN results with strong-field knowledge from the test-particle limit, and to further incorporate information from NR simulations. It consists of a Hamiltonian and a prescription for computing gravitational waves together with their back-reaction onto the orbit. While the EOB Hamiltonian describes generic motion, current state-of-the art refinements and calibrations to NR simulations~\cite{Bohe:2016gbl,Taracchini:2013rva,Damour:2001tu,Barausse:2009xi,Damour:2014sva} are specialized to circular orbits. The resulting prescription for computing complete inspiral-merger-ringdown waveforms for spinning binaries on circular orbits has been instrumental for extracting the science from LIGO's detections~\cite{Abbott:2016blz,Abbott:2016nmj, Abbott:2016izl, Abbott:2017vtc}. Considerations for extending the EOB approach to eccentric bound orbits have also been studied, such as instantaneous contributions to the radiation reaction forces for nonspinning eccentric binaries~\cite{Bini:2012ji} and spin couplings beyond the circular-orbit limit~\cite{Kavanagh:2017wot}. 

In this paper we specialize to nonspinning binaries for simplicity. Our approach to developing an eccentric EOB model employs a relativistic re-parameterization of the dynamical variables that is similar to the efficient description used for extreme mass-ratio inspirals, e.g.~\cite{Cutler:1994pb,Schmidt:2002qk,Drasco:2003ky}. The differences are that our treatment is specialized to nonspinning systems but applies for generic mass ratios. Instead of working with the EOB canonical coordinates that each change on all timescales, we divide the degrees of freedom of the binary into a set of phase variables and quantities that evolve only due to gravitational radiation. We discuss this treatment of the conservative dynamics both for general EOB potentials and when applying specific choices for these functions, including the re-summed potentials from Refs.~\cite{Bohe:2016gbl,Barausse:2009xi} that have been calibrated to results from numerical relativity simulations for quasi-circular inspirals. We make use of the fundamental frequencies of the motion to compute fluxes and waveforms, analogous to the strategy used to solve the Teukolsky equation in black-hole perturbation theory as described e.g. in Ref.~\cite{Drasco:2005is}. For transparency, we limit explicit results for quantities in the dissipative sector of the model to $1.5$PN order. This already requires calculating the instantaneous contributions in EOB instead of harmonic coordinates, and establishing a method for computing hereditary effects. Based on these theoretical tools, an extension to include all available PN information is straightforward but must be worked out carefully; this will be the subject of future work. We further demonstrate the procedure for building complete EOB waveforms that include inspiral, merger, and ringdown signals, by adapting the method for circular inspirals described in Ref.~\cite{Bohe:2016gbl}.

The methodology used here differs from the approach employed in PN calculations in the following ways. We use the Keplerian re-parameterization that applies for any eccentric orbit and is often used to describe geodesic motion around BHs. The description is based on the semilatus rectum $p$ and the eccentricity $e$, together with two phase variables associated with the spatial geometry of the radial and azimuthal motion denoted by $(\xi, \phi)$. 
These variables are defined by expressing the radial motion as
\be
r=\frac{p M}{1+e\cos\xi},
\ee
where $M$ is the total mass. The peri- and apoapsis correspond to $\xi=(0, \pi)\,{\rm mod \, 2\pi} $ respectively and define the parameters $(p,e)$. An important feature of the dynamics of a nonspinning relativistic binary on a bound eccentric orbit is that it is characterized by two frequencies: the radial frequency $\omega_r$ associated with the libration between the apo- and periapsis, and the azimuthal rotational frequency $\omega_\phi$. We introduce a
set of auxiliary phases $(\psi_r,\psi_\phi)$ associated with these frequencies defined by
\be
\frac{d\psi_r}{dt}=\omega_r(e,p), \ \ \ \ \ \ \ \frac{d\psi_\phi}{dt}=\omega_\phi(e,p).
\ee 
The utility of these phases is that they define the fundamental Fourier decomposition of quantities related to the dynamics; see also Refs.~\cite{Schmidt:2002qk,Hinderer:2008dm} for further discussion in a related context. For example, any function of $r$ can be expanded as $f(r)=\sum_{k} f_k(e,p) e^{ik\psi_r}$, where $k\in \mathbb{Z}$. We apply this decomposition to the binary's multipole moments from which the asymptotic gravitational radiation is computed as reviewed in Ref.~\cite{Blanchet2014}. The Newtonian mass quadrupole moment $I_{ij}$ has the form
\be
I_{ij}\sim \sum_{s=-\infty}^\infty\sum_{m=-2}^2 I^{ij}_{sm}(e,p)\, e^{i(s\psi_r+m\psi_\phi)}. \label{eq:Iijus}
\ee
Using the angles $(\psi_r,\psi_\phi)$ that directly reflect the frequency content of the underlying dynamics helps to make the calculations more transparent. Furthermore, it leads to an unambiguous definition of the orbit-average as the integral over one cycle of $\psi_r$, with $\psi_\phi$ remaining distinct.

By contrast, PN calculations generally rely on the quasi-Keplerian (QK) parameterization that leads to an explicit solution to the dynamics in terms of perturbative corrections to Newtonian motion, as reviewed e.g. in Refs~\cite{Arun:2007rg,Boetzel:2017zza}. This requires introducing several auxiliary variables: three ``eccentricities" ($e_t, e_r,e_\phi$), a semi-major axis ($a$), and three angle variables termed the true, eccentric, and mean anomaly, in addition to the azimuthal orbital phase $\phi$. The relative radial separation is expressed in terms of the eccentric anomaly $u$ as
\be
r^{\rm QK}=a(1-e_r\cos u). 
\ee
 The mean anomaly $\ell$ is the angle associated with the radial frequency  through $\ell=\omega_r t $, where $\omega_r$ is  referred to as the mean motion. The QK azimuthal phase is obtained in the form $\phi^{\rm QK}=(k+1)V(u)+f(u)$, where $k=(\omega_\phi/\omega_r-1)$ is the periastron advance that is assumed to be small, $k\ll 1$ as is appropriate for nearly Newtonian orbits, and the explicit expressions for $V$ and $f$ are not needed here. The binary's multipole moments are written as a Fourier decomposition of the form
\be
I_{ij}^{\rm QK}\sim \sum_{s=-\infty}^\infty\sum_{m=-2}^2 I^{ij \, {\rm QK}}_{sm}\, e^{i(s+mk)\ell}.\, \label{eq:IijQK}
\ee
This is similar to the decomposition in Eq.~\eqref{eq:Iijus} but with the notable difference that it involves only the radial phase variable $\ell$. For a fixed orbit, the form \eqref{eq:IijQK} is equivalent to (\ref{eq:Iijus}), however, the fact that $k$ is assumed to be non-integer complicates the calculations. Furthermore, because the dependence on $\psi_\phi$ has been recast in terms of $\ell$, the orbit-averaging operation that is defined as the integral over one cycle in $\ell$ becomes more involved, e.g. one must perform a series expansion for $k\to 0$ to evaluate averages~\cite{Arun:2007rg}. The assumption that $k\ll 1$ also makes it difficult to capture strong-field effects using this approach. On the other hand, an advantage of the QK parameterization is that the Fourier amplitudes $I^{ij \ {\rm QK}}_{sm}$ can be expressed in terms of Bessel functions that are convenient for asymptotic analyses. The parameterization used here does not immediately yield compact results in terms of known functions, however, our results can likewise be computed for arbitrary eccentricity. In addition, the numerical implementation of our approach requires regularizations, e.g. at turning points of the motion and for the circular-orbit limit. The regularizations can be accomplished by a similar treatment as developed in Ref.~\cite{Gair:2010iv} and as a result, these issues do not obstruct the practical use of the model, as we demonstrate in this work.

The organization of this paper is as follows. We start in Sec.~\ref{sec:reparam} by reviewing the EOB description of the binary's dynamics and work out its formulation in terms of the more efficient re-parameterization in terms of $(e,p,\xi,\phi)$. In Sec.~\ref{sec:frequencies} we discuss the fundamental properties of the conservative motion. Next, we focus on the dissipative sector and compute the fluxes and waveforms to 1.5PN order in Sec.~\ref{sec:fluxes}. We calculate instantaneous fluxes in EOB coordinates in Sec.~\ref{sec:instantfluxes}, then compute tail effects in Sec.~\ref{sec:tailfluxes}. For the hereditary effects we first calculate general results that involve the fundamental frequencies and Fourier amplitudes from the EOB dynamics, then specialize to the PN limit in order to derive explicit analytical results in a compact form. In Sec.~\ref{sec:fluxes} we also verify that our results for the orbit-averaged fluxes agree with known PN results when expressed in terms of gauge-invariant quantities. We compute the $h_{\ell m}$ modes in Sec.~\ref{sec:hlm}, where we also describe a procedure for attaching the merger-ringdown signals as a proof-of-principle that our approach enables computing complete waveforms. Subsequently, in Sec.~\ref{sec:implementation}, we address challenges in the practical implementation of our formalism and explain in detail how to overcome them. In Sec.~\ref{sec:results} we present several examples of waveforms and other quantities characterizing the binary computed in the adiabatic approximation and including only the limited PN information derived in Sec.~\ref{sec:hlm} in the waveform amplitudes. Section~\ref{sec:conclusion} contains our conclusions and the outlook on remaining tasks for future work. Finally, the Appendices contain details about the EOB potentials and the numerical treatment of the dynamics when using instantaneous radiation reaction forces. 

Throughout this paper we will use geometric units $G = 1 = c$.

\section{Re-parameterization of the conservative dynamics}
\label{sec:reparam}
\subsection{The effective-one-body Hamiltonian}
\label{sec:HEOB}
The EOB framework~\cite{Buonanno:1998gg} combines strong-field effects from the test-particle limit with finite mass-ratio corrections from the PN approximation. The model has additional flexibility to include nonperturbative information obtained from NR simulations. The conservative dynamics of the binary is described by the EOB Hamiltonian~\cite{Buonanno:1998gg}
\be
H_{\rm EOB}=M\sqrt{1+2\nu (\hat{H}_{\rm eff}-1)}. \,\label{eq:Hreal}
\ee
Here, $M=m_1+m_2$ is the total mass, $m_{1,2}$ are the masses of each object, and $\nu=m_1m_2/M$ is the symmetric mass ratio. The reduced effective Hamiltonian $\hat{H}_{\rm eff}=H_{\rm eff}/\mu$, where $\mu=\nu M$ is the reduced mass, describes an effective test-particle of mass $\mu$ moving in an effective metric on a non-geodesic trajectory. For nonspinning binaries moving in the plane $\theta=\pi/2$ the effective metric is that of a spherically-symmetric spacetime given by
\be
{\rm d}s^2_{\rm eff} = - A {\rm d}t^2 +\frac{{\rm d}r^2}{AD}+r^2 {\rm d}\phi^2. \label{eq:EOBmetric}
\ee
 The metric potentials $A$ and $D$ for the EOB model are given in the Appendix~\ref{sec:EOBpotentials}. They have the property that in the test-particle limit $\nu\to 0$ they reduce to the Schwarzschild potentials $A\to 1-2M/r$ and $D\to 1$. The effective Hamiltonian associated with the metric~\eqref{eq:EOBmetric} has the form~\cite{Damour:2000we}
\be
\hat{H}_{\rm eff}^2=A\bigg[1+\frac{P_\phi^2}{\mu^2 r^2}+\frac{ADP_r^2}{\mu^2}+\frac{Q_4(r)\, M^2\, P_r^4}{r^2\, \mu^4}+O(P_r^6)\bigg],\label{eq:Heffexpr} \ \ \ \ \ \ 
\ee
where $P_r$ and $P_\phi$ are the canonical radial and azimuthal angular momentum. The function $Q_4=2(4-3\nu)\nu+O(r^{-1})$ represents a non-geodesic term that appears at 3PN order; it is known to relative 1PN order where also the $O(P_r^6)$ contribution must be included. For simplicity, we will include only the 3PN nongeodesic term in our discussion; this can be extended in future work using the same methods as described here.

The energy of the system is given by
\be
{ E}=H_{\rm EOB},
\ee
which implies the relation
\be
\hat{H}_{\rm eff}(E)=1+\frac{1}{2\nu}\left(\frac{E^2}{M^2}-1\right). \label{eq:HeffofE}
\ee
Solving Eqs. (\ref{eq:Heffexpr}) and (\ref{eq:HeffofE}) for $P_r$ in terms of $(E,P_\phi,r)$ leads to
\be
\label{eq:prof}\hat{P}_r^2=\frac{2}{\beta D A}\left[-1+\sqrt{1+\beta Y}\right],
\ee
where we have defined the reduced momenta $\hat{P}_r=P_r/\mu$ and $\hat{P}_\phi=P_\phi/\mu$, and the coefficients are
\bea
\beta&=&\frac{4M^2 Q_4}{r^2 A^2 D^2}, \label{eq:betadef}\\
Y&=&\frac{\left(E^2+2\nu M^2-M^2\right)^2}{4\mu^2 M^2 A}-1-\frac{\hat{P}_\phi^2}{r^2}. \qquad\label{eq:Ydef}
\eea
In the test-particle limit of Eq.~\eqref{eq:prof},  $\beta\to 0$ and $\hat P_r$ becomes the radial momentum for a geodesic in Schwarzschild spacetime given by
\be
\lim_{\nu\to 0}\hat P_r^2=\frac{Y}{A D}+O(\beta).
\ee
The EOB equations of motion including radiation reaction can then be written as
\bes\label{eomsEOBrphiEL}\bea
\dot r&=&\frac{2A \mu M^2\, \hat{P}_r \left[r^2A D+2 M^2 Q_4\hat{P}_r^2\right]}{r^2E\left(E^2+2\nu M^2-M^2 \right)} , \; \; \; \; \; \; \; \; \; \label{eq:rdoteom}\\
\dot \phi&=&\frac{2\mu M^2 A\hat{P}_\phi}{r^2E\left(E^2+2\nu M^2-M^2\right)},\label{eq:phidotgen}\\
 \dot{E}&=& F_{\rm E}, \qquad  \qquad \dot{\hat P}_\phi= \hat{ F}_\phi,
\eea
\ees
where the substitution for $\hat P_r$ from Eq.~\eqref{eq:prof} is implied. 
The quantities $F_i$ are radiation reaction forces that are related to the gravitational wave fluxes of energy and angular momentum, as will be discussed in Sec.~\ref{sec:forces}.

Note that the equations of motion in the form given in Eqs.~\eqref{eomsEOBrphiEL} differ from the canonical EOB dynamics because the dependence on $\hat{P}_r$ has been eliminated in favor of $E$. This is already a more convenient formulation since unlike $\hat{P}_r$, the energy $E$ changes only due to radiation reaction, and the denominators in $\dot{x}^i$ are simple functions of $E$ instead of involving a complicated dependence on all EOB coordinates through $H_{\rm EOB}(x^i, P_i)$ as is the case in the canonical formulation. 

%
\subsection{Re-parameterization of the constants of motion}
\label{sec:constants}
The description of the dynamics in Eqs.~\eqref{eomsEOBrphiEL} can be further adapted to reflect the properties of the motion in the following way. For an eccentric bound orbit we define the Keplerian orbital elements $p$ and $e$ by
\be
r_{1}=\frac{pM}{1-e}, \ \ \ \ \ r_2=\frac{pM}{1+e}, \label{eq:r12}
\ee
where $r_{1,2}$ are the turning points of the radial motion. These turning points are computed by solving the radial equation of motion~\eqref{eq:rdoteom} for $\dot r=0, \, \hat{P}_r=0$. Setting to zero Eq.~\eqref{eq:prof} evaluated at $r_1$ and $r_2$ and solving for $(E,\hat{P}_\phi)$ in terms of $(p,e)$ gives
\bes
\label{eq:psofep}\bea
\hat{P}_\phi^2&=&\Bigg. \frac{p^2 M^2 \left(A(r_2)-A(r_1)\right)}{(1-e)^2A(r_1)-(1+e)^2A(r_2)}\Bigg. ,\\
\frac{ E^2}{M^2}&=&1-2\nu+\frac{4\nu \sqrt{e}\sqrt{A(r_1)A(r_2)}}{\sqrt{(1+e)^2A(r_2)-(1-e)^2A(r_1)}} \qquad \qquad
\eea
\ees
In the test particle limit $\nu \to 0$, the EOB parameters $(e,p)$ directly reduce to those for a geodesic in Schwarzschild spacetime, while the first integrals of the motion are related by $\{ \hat{P}_\phi, E\}\to \{ p_\phi^{\rm geo}, \, M+\mu E^{\rm geo}-\mu\}$. This can be seen by expanding $H_{\rm EOB}$ for $\nu\to 0$ and using that $H_{\rm eff}\to H^{\rm geo}$ in this limit. 
%
\subsection{Re-parameterization of the equations of motion}
\label{sec:eom}
We next introduce a phase variable $\xi$ defined by
\be
r=\frac{p M}{1+e\cos\xi} \label{eq:rofxi}
\ee
so that turning points of the motion correspond to $\xi=(0,\pi)\, {\rm mod}( 2\pi)$. Differentiating Eq.~(\ref{eq:rofxi}) leads to the following evolution equation
\be
\label{eq:xidotgen}
\dot \xi=\frac{(1+e\cos\xi)^2}{epM\sin\xi}\dot r+\frac{\cot\xi}{e}\dot e-\frac{1+e\cos\xi}{ep\sin\xi}\dot p.
\ee
The equations of motion for $(e,p)$ are obtained from those for $(E, P_\phi)$ by the transformation
\bes
\label{eq:fullset}
\be
\label{eq:edotpdot}
\dot{e}=c_{Ep}\frac{d\hat{P}_\phi}{dt}-c_{Lp}\frac{dE}{dt},\qquad
\dot{p}=c_{Le}\frac{dE}{dt}-c_{Ee}\frac{d\hat{P}_\phi}{dt}, \qquad
\ee
where the coefficients are given by
\be
\label{eq:cAbdef}
c_{Cb}=\frac{\partial C/\partial b}{(\partial E/\partial p)(\partial \hat{P}_\phi/\partial e)-(\partial E/\partial e)(\partial \hat{P}_\phi/\partial p)}.
\ee
Here, $C=\{E,\hat{P}_\phi\}$, and the derivatives are computed from the expressions in Eqs.~\eqref{eq:psofep}. 

The final set of EOB equations of motion are Eqs.~\eqref{eq:edotpdot} together with the evolution of the phases described by
\bea
\dot \xi&=& {\cal P}(e,p,\xi)+\frac{\cot\xi}{e}\dot e-\frac{1+e\cos\xi}{ep\sin\xi}\dot p, \; \; \; \; \; \; \; \; \label{eq:dotxi}\\
    \dot\phi&=&\Bigg. \frac{A \sqrt{A(r_1)-A(r_2)} (1+e \cos \xi)^2}{2 \sqrt{e} \, p \, E \sqrt{A(r_1)}
   \sqrt{A(r_2)} },\Bigg. 
    \eea
   with $E$ given in Eq.~\eqref{eq:psofep}. The function ${\cal P}$ governing the conservative dynamics of the radial phase variable $\xi$ is 
   \bea
  & &{\cal P}(e,p,\xi)= \Bigg. \frac{ A \hat{P}_r\sqrt{(1+e)^2 A(r_2)-(1-e)^2 A(r_1)}}{2 e^{3/2}p^3 \sin\xi \, E \sqrt{A(r_1)A(r_2)}}\\
   &&\qquad\quad\times (1+e \cos \xi )^2\left[p^2AD+2\hat{P}_r^2(1+e \cos \xi )^2  Q_4\right]. \Bigg. \nonumber \quad
    \eea
\ees
All the terms on the right hand sides have to be expressed in terms of only $(\xi, e,p)$ using Eqs.~(\ref{eq:prof}), (\ref{eq:psofep}), and (\ref{eq:rofxi}).

 \subsection{Fundamental properties of the conservative dynamics}
 \label{sec:frequencies}
Eccentric planar orbits possess two frequencies characterizing the radial librations between the turning points and the azimuthal rotations. In the Newtonian limit both of these frequencies coincide, however, this degeneracy is broken for relativistic motion. The frequencies are defined as follows. One period of the radial motion is the time elapsed between successive periapsis passages, and hence the time taken for $\xi$ to increase from $0$ to $2\pi$. From the conservative part of Eq.~(\ref{eq:dotxi}), the corresponding radial frequency is given by
 \be
 \omega_r=\frac{2\pi}{\int_0^{2\pi}d\xi/{\cal P}}. \label{eq:omegardef}
 \ee
 We associate to this frequency an angle variable $\psi_r$ defined by 
 \be
 d\psi_r/dt=\omega_r.
 \ee 
Any function of $r$ can thus be decomposed in a Fourier series as
 \be
 f(r)=\sum_{k=-\infty}^\infty {\rm f}_k e^{ik\psi_r},  \ \ \ \ \ \ \ {\rm f}_k=\frac{1}{2\pi}\int^{2\pi}_0 d \psi_r f e^{-ik\psi_r}. \ \ \ \ \label{eq:expandinqr}
 \ee
 The orbit-average $\langle f \rangle$ is the zero-coefficient computed from
 \be
 \langle f\rangle =\frac{1}{2\pi}\int^{2\pi}_0 d \psi_r f ={\rm f}_0.\ \ \ \ \label{eq:orbitavg}
 \ee
The relation between the two radial phase variables, $\psi_r$ associated with the frequency and $\xi$ with the orbit's spatial geometry is
  \be
  \label{eq:dqrdxi}
   \frac{d\psi_r}{d\xi}=\frac{\omega_r}{{\cal P}}. 
   \ee
Thus, integrals over $\psi_r$ can also be computed by converting them to integrals over $\xi$: 
\be
\label{eq:integralsqrtoxi}
\int d \psi_r f=\omega_r\int\frac{d\xi}{{\cal P}}f. 
\ee
It is often computationally more convenient to use the second expression in Eq.~\eqref{eq:integralsqrtoxi} since the relation $r(\xi)$ is rather simple and defined by ~\eqref{eq:rofxi} while the function $r(\psi_r)$ is given implicitly by a Fourier expansion as in Eq.~\eqref{eq:expandinqr}.
Defining the potential for the azimuthal motion to be the right-hand side of the equation of motion for $\phi$
\be
\dot \phi=V_\phi \label{eq:Vphidef}
\ee
we compute the azimuthal frequency from the orbit-average of the $\phi$ motion as
 \be
   \omega_\phi=\langle V_\phi\rangle=\frac{\int^{2\pi}_0\frac{d\xi}{{\cal P}}V_\phi}{\int^{2\pi}_0 d\xi/{\cal P}}=\frac{\omega_r}{2\pi}\int^{2\pi}_0\frac{d\xi}{{\cal P}}V_\phi. \qquad\label{eq:omegaphidef}
   \ee
For later use  in the Fourier expansion of the radiative multipole moments, we also note that the azimuthal phase $\phi$ can be decomposed into a linearly growing and an oscillatory part of the form
  \be
 \phi=\phi_0+\omega_\phi t+\Delta \phi_r, \label{eq:phiofxi}
 \ee
   where $\phi_0$ is an initial value and the oscillatory part is given by $\Delta \phi_r=\phi-\omega_\phi t$. It follows that the function $\Delta \phi_r$ can be expanded in a Fourier series as in Eq. (\ref{eq:expandinqr}). This can be seen explicitly by applying the decomposition~\eqref{eq:expandinqr} to Eq.~\eqref{eq:Vphidef} and using Eq.~\eqref{eq:orbitavg}:
   \be
   \dot \phi=\omega_\phi+\sum_{\substack{{k=-\infty}\\{k\neq 0}}}^{\infty} V_{\phi k}e^{i k\psi_r}. 
   \ee
Integrating both sides leads to
   \be
   \phi=\phi_0+\omega_\phi t+\sum_{\substack{{k=-\infty}\\{k\neq 0}}}^\infty \frac{V_{\phi k}}{ik \omega_r}e^{i k\psi_r},
   \ee
     where the last term is the oscillatory piece $\Delta\phi_r$ in Eq.~\eqref{eq:phiofxi}.
\section{Calculation of the fluxes and waveforms}
\label{sec:fluxes}

In this section we calculate the gravitational wave fluxes of energy and angular momentum, and the amplitude of the gravitational wave strain. In general, the fluxes can be obtained from the strain amplitudes, however, we do not consider this connection in this paper; it is an important subject of future work and necessary for a self-consistent model of the gravitational radiation and backreaction onto the dynamics. Instead, we first compute the fluxes and check our results against those from PN computations in the appropriate limit. Then we apply the same methods to calculate the spherical harmonic decomposition of the gravitational waveform. 
 
\subsection{Instantaneous contributions}
\label{sec:instantfluxes}

We first discuss the computation of the instantaneous piece of the fluxes of energy and angular momentum to 1PN order. This provides nontrivial checks of the method such as the transformation of the fluxes from harmonic or ADM coordinates to EOB gauge that can readily be extended to higher PN order. In all expressions, we will keep only terms up to 1PN order without indicating the omission of higher-order terms. We start from the PN results for the instantaneous parts of the fluxes in ADM or harmonic coordinates given by~\cite{Arun:2007sg}: 
\begin{widetext}
\bea
\label{eq:EFlux1PN}
{\cal F}_{1{\rm PN}}&=&\frac{32}{5} \frac{\mu^2M^2}{r^4} \bigg[v^2-\frac{11}{12} \dot{r}^2+\left(\frac{785}{336}-\frac{71}{28}\nu\right)v^4+\left(-\frac{1487}{168}+\frac{58}{7}\nu\right){\dot r}^2 v^2 +\left(-\frac{170}{21}+\frac{10}{21}\nu\right)\frac{M}{r} v^2\nonumber\\
&&+\left(\frac{687}{112}-\frac{155}{28}\nu\right)\dot r^4
+\left(\frac{367}{42}-\frac{5}{14}\nu\right)\frac{M}{r} {\dot r}^2 +\left(\frac{1}{21}-\frac{4}{21}\nu\right)\frac{M^2}{r^2} \bigg],\\
{\cal G}^z_{1{\rm PN}}&=& \frac{\mu^2 M\dot \phi}{r} \bigg[\frac{16}{5}v^2-\frac{24}{5} \dot{r}^2+\frac{16}{5}\frac{M}{r}+\left(\frac{614}{105}-\frac{1096}{105}\nu\right)v^4+\left(-\frac{296}{35}+\frac{1108}{35}\nu\right){\dot r}^2 v^2 +\left(-\frac{464}{105}-\frac{152}{21}\nu\right)\frac{M}{r} v^2\nonumber\\
&&+\left(\frac{38}{7}-\frac{144}{7}\nu\right)\dot r^4
+\left(\frac{496}{35}+\frac{788}{105}\nu\right)\frac{M}{r} {\dot r}^2 +\left(-\frac{596}{21}+\frac{8}{105}\nu\right)\frac{M^2}{r^2} \bigg],
\eea
\end{widetext}
where ${\cal F}$ denotes the energy flux and ${\cal G}^z$ the $z-$component of the angular momentum flux. 
Next, we use the transformation between EOB (denoted by a subscript E) and ADM (subscript A) coordinates given in Ref.~\cite{Bini:2012ji}:
\bes
\bea
x^i_{\rm A}&=&x^i_{\rm E}+\delta x^i_{\rm E},\label{eq:xiEOB}\\
\delta x^i_{\rm E}&=&\frac{\nu}{2}{\boldsymbol{\hat{P}}}_{\rm E}^2x_{\rm E}^i-\frac{(2+\nu)M\, x_{\rm E}^i}{2r_{\rm E}}+\nu \,r_{\rm E}\, \hat{P}^{E}_r\, \hat{P}_{\rm E}^i.  \; \; \; \; \qquad
\eea
\ees
The flux in EOB coordinates is then obtained from the expression in ADM coordinates~\eqref{eq:EFlux1PN} by the transformation 
\bes\label{eq:EOBF}
\bea
&&{\cal F}_{\rm EOB}={\cal F}_{\rm ADM}(r,\dot r,v^2)+\delta \mathcal{F}^{\rm A\,to\,E},\\
&&\delta \mathcal{F}^{\rm A\,to\,E}=\delta_{1, r} \frac{\partial F_{\rm ADM}}{\partial r}+\delta_{1, \dot r} \frac{\partial F_{\rm ADM}}{\partial \dot r}+\delta_{1, v^2} \frac{\partial F_{\rm ADM}}{\partial v^2}. \; \; \; \; \; \; \; \; \label{eq:deltaF}
\eea
\ees
Here, the notation is $r_{\rm A}=\sqrt{\delta_{ij}x^i_Ax^j_A}=r_{\rm E}+\delta_{1,r} $, where $\delta_{1,r}$ represents the correction at 1PN order, and similarly for $\dot r$ and $v^2$.
The corrections to $r$, $\dot r$ and $v^2$ are determined from Eq. (\ref{eq:xiEOB}) to be
\bes\label{eq:deltasHtoEOB}
\bea
\delta_{1,r}&=&\frac{x^i_E}{r_E} \delta x^i_E=-M+\nu\left(\frac{3}{2}r \dot r^2+\frac{1}{2}r^3\dot\phi^2-\frac{M}{2}\right), \; \; \; \; \; \; \; \; \; \\
\delta_{1,\dot r}&=&\frac{x^i_E}{r_E}\frac{d}{dt}\left(\delta x^i_E\right)+\frac{r_E\dot{x}^i_E-\dot r x^i_E}{r_E^2}\delta x^i_E\nonumber\\
&=&\nu\dot r\left(\frac{5}{2}v^2-{\dot r}^2-\frac{3M}{r}\right) \ \ \ \ \\
\delta_{1,v^2}&=&2 v^i_E \frac{d}{dt}\left(\delta x^i_E\right)\nonumber\\
&=&\frac{2M}{r} \left({\dot r}^2-v^2\right)+\nu\left(3v^4-\frac{3 (v^2+\dot r^2)M}{r}\right),
\eea
\ees
where the subscript $E$ has been omitted in the second equalities on the right hand sides. To obtain Eqs.~\eqref{eq:deltasHtoEOB} we have also used the relation 
\be
\hat P^i_{\rm E}=v^i_{\rm E}+O({\rm 1PN}) \label{eq:pofvEOB}
\ee
 and the EOB equations of motion expanded to 1PN order 
\be
\frac{d\hat{P}_i}{dt}=-\frac{x^iM}{r^3}+\frac{x^i M}{r^3}\left[\frac{(\nu-1)}{2}{\boldsymbol{v}}^2-\frac{(1+\nu)M}{r}-\dot{r}^2\right].\label{eq:dPdtEOB}
\ee
In Eq~\eqref{eq:dPdtEOB} we have used Eq.~\eqref{eq:pofvEOB} and omitted the subscript $E$ on the EOB coordinates, as we will continue to do in what follows below. 
Using Eqs.~\eqref{eq:deltasHtoEOB} in Eqs.~\eqref{eq:EOBF} gives for the additional contribution to the flux in EOB coordinates
\bea
\label{eq:deltaF}
\delta \mathcal{F}^{\rm A\,to\,E}&=&\frac{32\mu^2M^2}{15 r^5}\bigg[M\dot r^2(1-\nu)+3 Mr^2\dot \phi^2(2-\nu)\nonumber\\
&& \, \; \; \; \; \; \; -\frac{3}{4}\nu r{\dot r}^4-\frac{57}{4}\nu r^3\dot r^2\dot\phi^2+3 r^5\nu\dot\phi^4\bigg]. \; \; \; \; \; \; \; 
\eea
The angular momentum flux further involves the quantity $r^2\dot \phi=\epsilon_{3jk}x^j v^k$. Its  transformation is given by
\bea
\label{eq:phidotAtoE}
r_A^2\dot\phi_A&=&r_E^2 \dot{\phi}_E+\left(\dot y_E \delta x_E-\dot x_E \delta y_E+x_E\delta \dot y_E-y_E\delta \dot x_E\right)\nonumber\\
&=&r^2\dot \phi\left[1+2\left(\nu\, v^2-\frac{(1+\nu)M}{r}\right)\right].
\eea
Using these transformations leads to the following contribution to the angular momentum flux
\bea
\delta {\cal G}^z_{{\rm A\,to\,E}}&=& \frac{4\mu^2M\dot{\phi}}{5 r^3}\bigg[8M^2- Mr \dot r^2(2+3\nu)-\nu r^2 \dot r^4 \bigg.\qquad \; \; \; \\
&&\bigg.+14 r^6\nu \dot\phi^4-\left(4M +14 M\nu +17 \nu r \dot r^2\right)r^3\dot\phi^2\bigg].\nonumber
\eea
The next step is to substitute for $\dot r$ and $\dot \phi$ from the EOB dynamics. To compare with existing results we perform a 1PN expansion of the conservative EOB dynamics expressed in terms of $(e,p,\xi)$ in Eqs.~\eqref{eq:fullset}, which leads to
\bes\label{eq:eoms1PN}\bea
\dot r&=&\frac{eMp\sin\xi}{(1+e\cos\xi)^2}{\cal P},\\
{\cal P}&=& \frac{(1+e\cos\xi)^2}{Mp^{3/2}}\bigg[1-\frac{3(1+e\cos \xi)}{p}+\frac{\nu (1-e^2)}{2p}\bigg],\label{eq:calPPN}\qquad\\
\dot \phi&=& \frac{(1+e\cos\xi)^2}{Mp^{3/2}}\bigg[1-\frac{2e\cos \xi}{p}+\frac{\nu (1-e^2)}{2p}\bigg].\; \; \; \; \; 
\eea
\ees
Using these expansions in Eqs.~\eqref{eq:EFlux1PN} and~\eqref{eq:deltaF}, and keeping only terms at 1PN order, gives an expression for the instantaneous energy flux. Since this step involves only straightforward substitutions we do not write out the results explicitly here. 

Since the instantaneous fluxes are gauge-dependent, it is easier to compare results for the orbit-averaged fluxes between different approaches. The average is computed from Eq.~\eqref{eq:orbitavg} and making use of the conversion from integrals over $\psi_r$ to integrals over $\xi$ from Eq. \eqref{eq:integralsqrtoxi}. The radial frequency, calculated from Eqs.~\eqref{eq:omegardef} and \eqref{eq:calPPN}, is given by
\be
\label{eq:omegarPN}
M\omega_r=\frac{(1-e^2)^{3/2}}{p^{3/2}}\left[1+\frac{(1-e^2)(-6+\nu)}{2p}\right].
\ee
Using this result, performing the averages of the fluxes, and truncating at 1PN order leads to
\bea
\label{eq:Fresultep}
\langle{\cal F}\rangle &=&\frac{32\mu^2(1-e^2)^{3/2}}{5 p^5M^2}\bigg\{1+\frac{73}{24}e^2+\frac{37}{96}e^4\\
&&+\frac{1}{p}\bigg[-\frac{1247}{336}-\frac{5\nu}{4}-e^2\left(\frac{9181}{672}+\frac{325\nu}{24}\right)\nonumber\\
&&+e^4\left(\frac{809}{128}-\frac{435\nu}{32}\right)+e^6\left(\frac{8609}{5376}-\frac{185\nu}{192}\right)\bigg]\bigg\},\nonumber\\
\langle {\cal G}^z\rangle&=&\frac{32\left(1-e^2\right)^{3/2}\mu^2}{5M p^{7/2}}\bigg\{1+\frac{7}{8}e^2\nonumber\\
&&+\frac{1}{p}\left[-\frac{1247}{336}-\frac{7\nu}{4}-e^2\left(\frac{425}{336}+\frac{401\nu}{48}\right)\right.\nonumber\\
&&\left.+e^4\left(\frac{10751}{2688}-\frac{205\nu}{96}\right)\right]\bigg\}.
\eea
The expression (\ref{eq:Fresultep}) is written in terms of the quantities $(e,p)$ that are defined by the EOB dynamics and therefore gauge dependent. Thus, we next express it in terms of less gauge-dependent quantities such as the energy and angular momentum or quantities related to the frequencies. Convenient quantities to consider for this purpose are
\bes
\label{eq:xiotaepsj}
\bea
x&=&(M\omega_\phi)^{2/3}, \label{eq:xiotadef}\\
\epsilon&=& -\frac{2(E-M)}{\mu}, \ \ \ \ \ \ j=-\frac{2(E-M)\hat{P}_\phi^2}{\mu M^2}. \label{eq:epsjdef} \qquad \quad
\eea
\ees
Existing PN results are usually given in terms of $(x,e_t)$, where $e_t$ is one of the eccentricities in the QK parameterization. To convert between $e_t$ and the EOB eccentricity $e$ it is easiest to proceed as follows. First, we compute the quantities from Eq.~\eqref{eq:xiotaepsj} in terms of the EOB parameters. The PN limit of the azimuthal EOB frequency is
\be
\label{eq:omegaphiPN}
M\omega_\phi=\frac{(1-e^2)^{3/2}}{p^{3/2}}\left[1+\frac{\nu+e^2(6-\nu)}{2p}\right].
\ee
A check on these results is that the test-particle $(\nu=0)$ limit of Eqs. (\ref{eq:omegarPN}) and \eqref{eq:omegaphiPN} agrees with Eqs. (5.1) and (5.2) of Ref.~\cite{Forseth:2015oua}. Inserting
the PN limit of the relations from
Eq.~\eqref{eq:psofep} into the definitions \eqref{eq:epsjdef}
leads to
\bes
\label{eq:epsjofep}
\bea
\epsilon&=&\frac{(1-e^2)}{p}\left[1+\frac{(1-e^2)(\nu-3)}{4p}\right],\label{eq:epsofe}\\
j&=& (1-e^2)\left[1+\frac{9+\nu+e^2(7-\nu)}{4p}\right].\label{eq:jofe}
\eea
\ees
Using Eq.~\eqref{eq:omegaphiPN} in Eq.~\eqref{eq:xiotadef} gives an expression for $x(e,p)$, which can be inverted to obtain
\be
p=\frac{1-e^2}{x}+\frac{1}{3}\left[\nu+e^2\left(6-\nu\right)\right]. \label{eq:pofx}
\ee
The PN parameter $e_t$ is given in terms of $\epsilon$ and $j$, e.g., in Eq.~(7.7e) of Ref.~\cite{Arun:2007sg}. Inserting therein the EOB variables from~\eqref{eq:epsofe} and~\eqref{eq:jofe} we obtain the following relation between the eccentricity parameters
\be
e_t^2=e^2\left[1+\frac{2}{p}(1-e^2)(\nu-3)\right]. \label{eq:etofe}
\ee

Using the relations~\eqref{eq:pofx} and ~\eqref{eq:etofe} to transform the fluxes to the PN variables leads to agreement with Eqs. (8.8)--(8.9b) of Ref.~\cite{Arun:2007sg} for the energy flux, and with Eqs. (4.10) and (4.11b) of Ref.~\cite{Arun:2009mc} for the angular momentum flux. 
\subsection{Hereditary contributions}
\label{sec:tailfluxes}

We next discuss the treatment of hereditary contributions. Our strategy closely follows the treatment used in black hole perturbation theory, and specifically the detailed discussion of this method provided in Ref.~\cite{Drasco:2005is}. We apply the Fourier expansion discussed in Sec.~\ref{sec:frequencies} to the radiative multipole moments and explain how this approach simplifies the computations compared to calculations based on the QK parameterizations that also employ a Fourier decomposition but use only the radial phase variable. We first derive general results that apply for fully relativistic dynamics and arbitrary eccentricity. These can be evaluated numerically in an EOB evolution. Next, we specialize to the PN limit of the dynamics to obtain explicit analytical expressions for the orbit-averaged fluxes and verify that, upon further specializing to low eccentricity, our results are in agreement with known PN expressions from Refs.~\cite{Arun:2007rg,Arun:2009mc}. 
\subsubsection{Fourier expansion of the radiative multipole moments}
The Newtonian mass multipole moments of the binary are given by
\be
I_L=\mu r^\ell n^L, \label{eq:IijN}
\ee
where $n^i=x^i/r$ are unit vectors, and $L$ denotes a string of spatial indices on tensors that are symmetric and trace-free. The unit vectors are related to spherical harmonics $Y_{\ell m}$ by
\be
Y_{\ell m}={\cal Y}_L^{\ell m}n_L ,
\ee
where ${\cal Y}_L^{\ell m}$ are constant tensors. They satisfy the identity 
\be
{\cal Y}_L^{\ell m \, *}{\cal Y}_L^{\ell m^\prime}=\frac{(2\ell+1)!!}{4\pi \ell !}\delta_{m m^\prime}, \label{eq:calYid}
\ee
where the summation on repeated indices is implied. 
The multipole moments in Eq.~\eqref{eq:IijN} can thus also be expanded as
\bea
I^{L}&=&\frac{4\pi \ell!}{(2\ell+1)!!}\mu r^\ell \sum_{m=-\ell}^\ell {\cal Y}^{L}_{\ell m} Y_{\ell m}^*(\theta, \phi)\nonumber\\ &=& \sum_{m=-\ell}^\ell {\cal Y}^{L}_{\ell m}a_{\ell m} r^\ell e^{-im\phi}. \label{eq:ILrewrite}
\eea
Here, we have specialized to $\theta=\pi/2$, and the coefficients $a_m$ are given by
\be
a_{\ell m}=\frac{4 \pi\ell! \mu}{(2\ell+1)!!}Y_{\ell m}^*\left(\frac{\pi}{2}, 0\right). \label{eq:almdef}
\ee

From Eq.~(\ref{eq:phiofxi}) it follows that the decomposition (\ref{eq:ILrewrite}) can be expressed as
\be
I^{L}=\sum_{m=-\ell}^\ell a_{\ell m} {\cal Y}_L^{\ell m} J_{\ell m}e^{-im \psi_\phi}. \label{eq:iLinter}
\ee
Here, the functions $J_{\ell m}$ are defined by 
\be
J_{\ell m}=r^\ell e^{-im\phi_0}e^{-im\Delta\phi_r}=\sum_{k=-\infty}^\infty J_{\ell mk}e^{-ik\psi_r}, \label{eq:Jlmdef}
\ee
where 
\bes
\bea
J_{\ell mk}&=&\frac{1}{2\pi}\int_0^{2\pi}d \psi_r e^{ik\psi_r}J_{\ell m}\\
&=&\frac{\omega_r}{2\pi}\int_0^{2\pi}\frac{d\xi}{{\cal P}}r^\ell e^{-im\phi_0}e^{-im\Delta\phi_r}e^{ik\psi_r} \ \ \label{eq:Jlmkgeneral}
\eea
\ees
Using Eq.~\eqref{eq:Jlmdef} in Eq.~\eqref{eq:iLinter} leads to the final Fourier decomposition of the Newtonian mass multipole moments
\be
I^L=\sum_{m=-\ell}^\ell\sum_{k=-\infty}^\infty {\cal Y}_L^{\ell m} a_{\ell m}  J_{\ell mk}e^{-i(k\psi_r+m\psi_\phi) }, \label{eq:IFourier}
\ee
which makes the biperiodic structure manifest. A similar decomposition applies for the current moments and PN corrections to the multipoles, however, they are not needed for the $1.5$PN tail terms considered here. An advantage of this parameterization compared to the QK parameterization is that the Fourier decomposition \eqref{eq:IFourier} is explicitly a function of two angular variables that are independent. By contrast, in the QK parameterization as summarized e.g. in Ref.~\cite{Arun:2007rg}, the phase variable $\psi_\phi$ is generally not used. Instead, it is eliminated by using the fact that for a fixed orbit $\psi_i=\omega_i t$ which implies that $\psi_\phi=\psi_r(\omega_\phi/\omega_r)$ in this case. We will discuss the disadvantages of this replacement below. 

The computations of the hereditary effects involve the $n$th time derivative of the multipole moments. For conservative dynamics it is given by
   \be
   I_L^{(n)}=\sum_{m=-\ell}^\ell\sum_{k=-\infty}^\infty (-i)^n\ \Omega_{mk}^n\ {\cal Y}_L^{\ell m} a_{\ell m}  J_{\ell mk}e^{-i(k\psi_r+m\psi_\phi) } \label{eq:dIFourier},  \ \ \ \ 
   \ee
   where we have defined the combination of frequencies
 \be
  \Omega_{mk}=m\omega_\phi+k\omega_r .  \label{eq:omegamk}
  \ee
  This can also be generalized to include the evolution of $(e,p)$ in future work. 
  
\subsubsection{Tail terms in the fluxes at 1.5PN order}
The $1.5$PN tail terms in the energy and angular momentum fluxes are~\cite{Blanchet2014}
\bes
\label{tails15gen}
\bea
{\cal F}_{{\rm tail},1.5{\rm PN}}&=& \frac{4M}{5}I_{ij}^{(3)}\int_0^\infty d\tau I_{ij}^{(5)}(t-\tau)\ln\left(\frac{\tau}{b}\right), \ \ \ \ \ \ \ \label{eq:Ftailgen15}\qquad\\
{\cal G}_{{\rm tail},1.5{\rm PN}}^k &=&  \frac{4M}{5}\epsilon_{kij}\bigg[I_{in}^{(2)}\int_0^\infty d\tau I_{jn}^{(5)}(t-\tau)\ln\left(\frac{\tau}{b}\right)\nonumber\\
&& \; \; \; \; +I_{jn}^{(3)}\int_0^\infty d\tau I_{in}^{(4)}(t-\tau)\ln\left(\frac{\tau}{b}\right)\bigg], \ \ \ \ \ \ \ \label{eq:Gtailgen15}
\eea
\ees
where $b=2 r_0e^{-11/12}$ and $r_0$ is a PN gauge parameter. The coefficients in Eq.~(\ref{eq:IFourier}) for $\ell=2$ are
    \be
   a_{20}=-\frac{2\mu}{3}\sqrt{\frac{\pi}{5}},  \ \ \ \ \ a_{22}=a_{2-2}=\mu\sqrt{\frac{2\pi}{15}}, \ \ \ \ \ \label{eq:acoeffs}
   \ee
   and $a_{21}=0=a_{2-1}$. 
   
  We first discuss a simplification of the structure of the tail fluxes, assuming that the binary is on a fixed orbit. Using Eq.~\eqref{eq:dIFourier} in Eqs.~\eqref{tails15gen} shows that they require evaluating terms of the general form (needed here with $n=3$ and $s=5$ but we will keep the discussion more general)
  \begin{widetext}
  \bea
  {\cal F}&=&\frac{4M}{5}I_L^{(n)}\int_0^\infty d\tau I_{L}^{(s)}(t-\tau)\ln\left(\frac{\tau}{b}\right)\nonumber\\
 &=&\frac{4M}{5}\sum_{m=-\ell}^\ell\sum_{m^\prime=-\ell}^\ell \sum_{ k=-\infty}^\infty \sum_{k^\prime=-\infty}^\infty (-i)^{n+s}\ \Omega_{mk}^n\ \Omega_{-m^\prime k^\prime}^s \ a_{\ell m}\ a_{\ell -m^\prime}\ J_{\ell m k}\ J_{l -m^\prime k^\prime}{\cal Y}_L^{\ell m} \ (-1)^{m^\prime}\ {\cal Y}_L^{\ell m^\prime *} \nonumber\\
 && \; \; \; \; \; \; \; \; \; \; \; e^{-i (k+k^\prime)\psi_r-i(m-m^\prime)\psi_\phi}\int_0^\infty d\tau e^{i\Omega_{-m^\prime k^\prime}\tau} \ln\left(\frac{\tau}{b}\right), \label{eq:calFmess}
 \eea
 where we have relabeled $m^\prime \to - m^\prime$ and used the identity ${\cal Y}_L^{\ell - m}=(-1)^m{\cal Y}_L^{\ell m  *} $. We have also used that for a fixed orbit $\psi_i(\tau)=\omega_i\tau$ and the definition \eqref{eq:omegamk}. This form of the expression enables us to use the orthogonality relation (\ref{eq:calYid}) and reduce Eq.~(\ref{eq:calFmess}) to 
 \be
 \label{eq:calFsimp}
 {\cal F}=\frac{4M}{5} \frac{(2\ell+1)!!}{4\pi \ell !} \sum_{m=-\ell}^\ell\sum_{\substack{ k=-\infty}}^\infty \sum_{\substack{ k^\prime=-\infty}}^\infty (-1)^m(-i)^{n+s}\ \Omega_{mk}^n\ \Omega_{-m k^\prime}^s a_{\ell m}^2 \ J_{\ell m k}J_{l -m k^\prime} e^{-i (k+k^\prime)\psi_r}\int_0^\infty d\tau e^{i\Omega_{-m k^\prime}\tau} \ln\left(\frac{\tau}{b}\right), \qquad
 \ee
 \end{widetext}
 where we have used that $a_{\ell m}=a_{\ell -m}$, which follows from the definition (\ref{eq:almdef}). 
In general, this is a simpler expression than that usually obtained from the QK analyses. Except in the special case of  Newtonian binary dynamics, the QK results for the fluxes still involve four summations, as can be seen e.g. in Eq. (4.20) in Ref.~\cite{Arun:2007rg}, where a dependence of the form $\sim {\rm exp}\left[{i (s+s^\prime+(m+m^\prime)k)\ell}\right]$ remains, with the variable $\ell$ being analogous to $\psi_r$ and $k$ being the periastron advance not an integer. By contrast, in the parameterization employed in Eq.~(\ref{eq:calFsimp}) the dependence on $\psi_\phi$ has been eliminated automatically from the orthogonality properties of the ${\cal Y}_{L}^{\ell m}$ tensors. A consequence of the residual factors in the QK approach is that evaluating the orbital average of the fluxes requires a series expansion for $k\ll 1$. From Eq.~(\ref{eq:calFsimp}) and the definition of the averaging operation \eqref{eq:orbitavg} it follows that within the more transparent decomposition employed here the orbit averaged flux does not require any approximations. 

 We next substitute the decomposition~(\ref{eq:dIFourier}) specialized to $\ell=2$ into Eq.~(\ref{eq:Ftailgen15}) and introduce the notation for the definite integral
 \bes
\bea
   {\cal I}(x)&=& \Bigg.  \int_0^\infty d\tau e^{i x\tau}\ln\left(\frac{\tau}{b}\right) \Bigg. \\
   &=&\Bigg. -\frac{1}{x}\bigg[\frac{\pi}{2}{\rm sgn}(x)+i{\rm ln} (|x|b)+i\gamma_E\bigg], \Bigg. \label{calIdef} \ \ \ \ \ \ \ 
 \eea
 \ees
 where $\gamma_E$ is the Euler constant. 
Splitting the Fourier expansion into the orbit-averaged and oscillatory pieces then leads to the following expression for the energy flux:
 \begin{widetext}
 \bes
 \label{eq:FGtailinst}
  \bea
  \label{eq:Ftailinst}
{\cal F}_{\rm tail}&=& 192Ma_{22}^2J_{220}^2\,\omega_\phi^7+\frac{3M}{2}\sum_{m=-2}^2\sum_{ k=1}^\infty a_{2m}^2J_{2mk}^2|\Omega_{mk}|^7\nonumber\\
&&+\frac{3Ma_{22}^2}{2\pi}\sum_{m\neq0}\sum_{\substack{ k=-\infty\\ k\neq 0}}^\infty e^{-ik\psi_r}J_{220}\,J_{2-mk}\,\Omega_{-mk}^3 \Omega_{m0}^3\left[\Omega_{-mk}^2{\cal I}(\Omega_{-mk})+\Omega_{m0}^2{\cal I}(\Omega_{m0})\right]\nonumber\\
&&+ \frac{3M}{2\pi}\sum_{m=-2}^2\sum_{\substack{ k=-\infty\\ k\neq 0}}^\infty \sum_{\substack{ k^\prime=-\infty\\ k^\prime\neq 0, -k}}^\infty e^{-i(k+k^\prime)\psi_r}a_{2m}^2J_{2mk}\,J_{2-m k^\prime}\,\Omega_{mk}^3\,\Omega_{-mk^\prime}^5\,{\cal I}(\Omega_{-mk^\prime}). 
 \eea
Here, the terms in the first line are the non-oscillatory contributions. To express them in this form we have used that ${\cal I}(x)+{\cal I}(-x)=-\pi /|x|$ and $J_{220}=J_{2-20}$ due to symmetry and the quadrupole being real. In the second term of the first line, which comes from the $k^\prime=-k$ contribution of the double sum, we have also rewritten the sum to be only over positive $k$, used that $J_{2mk}=J_{2-m-k}$ and $J_{2-m k}=J_{2m-k}$, and the freedom to re-label $m\to -m$ since $m$ is summed over the same negative and positive integers. Note that in cases where for some integers $\Omega_{sn}=0$, e.g. for a Newtonian orbit or cases with resonances, the corresponding terms in Eq. (\ref{eq:Ftailinst}) will give a vanishing contribution even though ${\cal I}(0)$ diverges, as can be seen from the original expressions in Eqs.~\eqref{eq:dIFourier} and \eqref{eq:Ftailgen15}.
  
  Similarly, for the angular momentum flux we obtain at 1.5PN order
 \bea
 \label{eq:Gtailinst}
{\cal G}_{\rm tail}^z&=& 192Ma_{22}^2J_{220}^2\, \omega_\phi^6 +\frac{3Ma_{22}^2}{\pi}\sum_{m\neq0}\sum_{\substack{ k=-\infty\\ k\neq 0}}^\infty  {\rm sgn}(m)\,J_{2-m k}^2\Omega_{-mk}^7{\cal I}(\Omega_{m-k})\nonumber\\
&&+\frac{3Ma_{22}^2}{2\pi}\sum_{m\neq 0}\sum_{\substack{ k=-\infty\\ k\neq 0, k\neq -2}}^\infty \, e^{-ik\psi_r} \, {\rm sgn}(m)\, J_{2m0}\, J_{2-mk}\,\Omega_{m0}^2\,\Omega_{-mk}^2\, \Omega_{(-2m)k}\, \left[\Omega_{-mk}^2 {\cal I}(\Omega_{-mk})+\Omega_{-m0}^2 {\cal I}(\Omega_{m0})\right]\nonumber\\
&&+\frac{3Ma_{22}^2}{2\pi}\sum_{m\neq0}\sum_{\substack{ k=-\infty\\ k\neq 0}}^\infty \sum_{\substack{ k^\prime=-\infty\\ k^\prime \neq 0,-k}}^\infty e^{-i(k+k^\prime)\psi_r}\, {\rm sgn}(m)\,J_{2-m k}\, J_{2mk^\prime}\Omega_{(-2m)(k-k^\prime)}\Omega_{-mk}^2\Omega_{mk^\prime}^4{\cal I}(\Omega_{mk^{\prime}}).
\eea
\ees
\end{widetext}
The terms in the first line are the non-oscillatory contributions that have been separated out from the remaining terms. 

As mentioned above, the orbit-averaged flux is readily obtained from Eqs.~(\ref{eq:FGtailinst}) by noting that the average, defined in Eq. (\ref{eq:orbitavg}), is only nonvanishing when the phase of the exponentials is zero. This implies that for the terms in Eq.~(\ref{eq:Ftailinst}) that involve only the factor of $e^{ik\psi_r}$ vanish. Likewise, since the averaging of the terms involving $ e^{i(k+k^\prime)\psi_r}$ produces a factor of $\delta_{(k+k^\prime),0}$ but $k^\prime=-k$ is excluded from the summation, those terms also vanish. The averaged 1.5PN tail fluxes thus reduce to the compact form
\bes
\label{eq:averagedfluxes}
\bea
 \langle {\cal F}_{\rm tail}\rangle&=& \frac{3M}{2}\bigg[a_{22}^2J_{220}^2\Omega_{20}^7\nonumber\\
 &&\qquad+\sum_{\substack{ k=1}}^\infty\sum_{m=-2}^2 a_{2m}^2J_{2mk}^2\, |\Omega_{mk}|^7 \bigg],\\
 \langle {\cal G}^z_{\rm tail}\rangle&=& 3M\,a_{22}^2 \bigg[ \, \Omega_{20}^6\, J_{220}^2\label{eq:Gavg}\\
 && \; \; +\sum_{\substack{ k=1}}^\infty \left(\Omega_{2k}^6\, J_{22k}^2-J_{2-2k}^2\, \Omega_{-2k}^6\,  {\rm sgn}(  \Omega_{-2k})\right)\bigg] . \ \ \ \nonumber
 \eea
 \ees
To obtain the expression~\eqref{eq:Gavg} from the first line of Eq.~\eqref{eq:Gtailinst} we have re-written the sum to be only over positive values of $k$,  explicitly performed the summation over $m$, and used the identity ${\cal I}(x)+{\cal I}(-x)=-\pi /|x|$. 
The expressions~\eqref{eq:averagedfluxes}, or their non-averaged  counterparts from Eqs.~\eqref{eq:FGtailinst}, can be evaluated numerically for a relativistic EOB trajectory with arbitrary eccentricity. 

\subsubsection{Explicit Fourier coefficients for post-Newtonian conservative dynamics}
\label{eq:Newtexplicit}
To check the results for the fluxes from Eq.~\eqref{eq:averagedfluxes} and make explicit their dependence on the parameters requires further approximations to the trajectory on which the coefficients and frequencies are computed. 
Specializing to the PN limit of the EOB model, the 1.5PN tails only require the Newtonian conservative dynamics from the leading order terms in Eqs.~(\ref{eq:eoms1PN}), \eqref{eq:omegarPN}, and \eqref{eq:omegaphiPN}. 
From Eqs.~(\ref{eq:eoms1PN}) truncated at Newtonian order we also obtain the relations \be
\phi=\xi, \ \ \ \  \Delta \phi_r=\phi-\omega_\phi t=\xi-\psi_r, \ \ \ \ \ \ \ \ \ \ \ \ \ \ \label{eq:deltaphirN}
\ee
since $\omega_\phi=\omega_r$ in this limit and $\psi_r=\omega_r t$ for the conservative dynamics. The relation between the variables $\psi_r$ and $\xi$ is found by integrating Eq. (\ref{eq:dqrdxi}) using the Newtonian limit of ${\cal P}$ from (\ref{eq:eoms1PN}). The result is  
\be \label{eq:qrofxiNewt}
\psi_r(\xi)=2\tan^{-1}\bigg[\frac{\sqrt{1-e}\tan\left(\frac{\xi}{2}\right)}{\sqrt{1+e}}\bigg]-\frac{e\sqrt{1-e^2}\sin\xi}{1+e\cos\xi},
\ee
where we omitted any integration constants since only trigonometric functions of $\xi$ will be needed. 
The coefficients $J_{\ell mk}$ are given by
   \bea
        J_{2mk}^{\rm Newt}&=&\frac{1}{2\pi}\int_0^{2\pi}r^2 e^{-ik\psi_r}e^{im\Delta\phi_r}d\psi_r    \label{eq:Jmkcoeffsgen}\\
        &=&M^3p^{7/2}\frac{\omega_r}{2\pi}\int_0^{2\pi}\frac{d\xi}{(1+e\cos\xi)^4}e^{im\xi}e^{-i(k+m)\psi_r(\xi)}. \nonumber \; \; \; 
    \eea
Here, we have set the initial phase $\phi_0=0$, used the symmetries $J_{2m-k}=J_{2-mk}$ to obtain Eq.~\eqref{eq:Jmkcoeffsgen} from Eq.~\eqref{eq:Jlmkgeneral}, and in the second line substituted ${\cal P}$ in the Newtonian limit from Eq.~\eqref{eq:eoms1PN}. For further analysis of the coefficients it is useful to express them using binomial expansions. The exponential involving $\psi_r$, using the result from Eq.~\eqref{eq:qrofxiNewt}, can then be written as the following expansion
 \begin{widetext}
\bea
e^{-ik\psi_r}&=&e^{ik \frac{e\sqrt{1-e^2}\sin\xi}{1+e\cos\xi}}\left(\frac{1+\sqrt{1-e^2}+e \, e^{i\xi}}{e+\left(1+\sqrt{1-e^2}\right)e^{i\xi}}\right)^k\nonumber\\
&=&\sum_{n=0}^\infty\sum_{s=0}^{\infty}\sum_{\ell=0}^\infty \frac{k^s}{s!}\begin{pmatrix} k\\ n\end{pmatrix}\begin{pmatrix} -k\\ \ell\end{pmatrix} i^s \, e^{s+\ell+n}\left(1+\sqrt{1-e^2}\right)^{-\ell-n}e^{-i\xi(k+\ell-n)}\left(\sin\xi\right)^s\left(1+e\cos\xi\right)^{-s},\label{eq:exppsirexp}
\eea
where $(:)$ are generalized binomial coefficients. For $k>0$ the sum over $n$ terminates at $n=k$, while for $k<0$, the sum over $\ell$ terminates at $\ell=k$. 
Using this expansion in Eq.~\eqref{eq:Jmkcoeffsgen}, converting the trigonometric functions to exponentials, and performing further binomial expansions, leads to 
\bes\label{eq:Jmks}\bea
J_{2mk}&=&\frac{M^3\omega_r p^{7/2}}{2\pi}\sum_{n=0}^{\infty}\sum_{\ell=0}^\infty\sum_{s=0}^\infty\sum_{w=0}^{s}\sum_{z=0}^\infty\sum_{v=0}^z \frac{(k+m)^s}{2^{z+s}s!}(-1)^{w+s}\,\begin{pmatrix} k+m\\ n\end{pmatrix}\begin{pmatrix} -(k+m)\\ \ell\end{pmatrix}\begin{pmatrix} s\\ w\end{pmatrix}\begin{pmatrix} z\\ v\end{pmatrix}\begin{pmatrix} -4-s\\ z\end{pmatrix} \nonumber\\
&& \times \, e^{s+\ell+n+z}\left(1+\sqrt{1-e^2}\right)^{-\ell-n}(1-e^2)^{s/2}\, \oint d\xi \, e^{-i\xi\left(k+\ell+z+s-n-2w-2v\right)}.
\eea
\ees
\end{widetext}
The integral is readily evaluated to be \mbox{$2\pi \delta_{\left(k+\ell+z+s-n-2w-2v\right),0}$}. 

To compare with existing PN results we specialize Eqs.\eqref{eq:Jmks} to the limit $e\ll 1$. The results truncated to $O(e^4)$ are
 \bes\label{eq:Jmksapprox}\bea
J_{20k}&=& -\frac{ M^2p^2}{4}\bigg[\frac{e^4}{3}\delta_{|k|,4}+\frac{e^3}{2}\delta_{|k|,3}\\
&&+\left(e^2+\frac{5e^4}{3}\right)\delta_{|k|,2}+\left(4e+\frac{15e^3}{2}\right)\delta_{|k|,1}\bigg], \ \nonumber\\
J_{2m0}&=&M^2p^2\left(1-\frac{1}{2}e^2-\frac{9}{16}e^4\right), \ \ \ \ \ \ \\
J_{2mk}&= &\frac{ M^2p^2}{8}\bigg\{\frac{e^4}{8}\left(34\pm 19m\right)\delta_{k,\pm 4}+\frac{e^3}{3}\left(9\pm 8m\right)\delta_{k,\pm 3}\nonumber\\
&&   +\left[e^2(14\mp 3m)+e^4(18\mp11m)\right]\delta_{k,\pm 2}\nonumber\\
&&  +\left[8e(\pm m-1)+e^3(\pm 8m-19)\right]\delta_{k,\pm 1}\bigg\},  \ \ \ \ \; \; \; \;  \ \ \ \ 
 \eea
 \ees
where $J_{2m0}=J_{220}=J_{2-20}$ and the coefficients $a_m$ were given in Eq.~(\ref{eq:acoeffs}).

Finally, substituting these results \eqref{eq:Jmksapprox} into the expressions for the fluxes from Eqs.~\eqref{eq:averagedfluxes}, performing the sums, using the Newtonian relations $\omega_r=\omega_\phi$, $p=(1-e^2)/x$, and the definition $M\omega_\phi=x^{3/2}$, and re-expanding to $O(e^4)$ leads to
\bes
\label{eq:Ftailapproxresult}
\bea
 \langle {\cal F}_{\rm tail}\rangle &=&\frac{32\nu^2x^5}{5} 4\pi x^{3/2}\left[1+\frac{2335 e^2}{192}+\frac{42955 e^4}{768}\right], \; \; \; \; \; \; \; \; \; \; \\
 \langle {\cal G}^z_{\rm tail}\rangle &=&\frac{32M\nu^2x^{7/2}}{5}4\pi x^{3/2}\left[1+\frac{209 e^2}{32}+\frac{2415 e^4}{128}\right].\qquad
 \eea
\ees
From the mapping to the PN eccentricity parameter from Eq.~\eqref{eq:etofe} we see that at the order needed here $e=e_t$ so that the above expressions for the fluxes can be directly compared with PN calculations. The energy flux of Eq.~\eqref{eq:Ftailapproxresult} is in agreement with the result recalled in Eq. (7.1a) of Ref.~\cite{Arun:2009mc} and also computed e.g. in Eq.~(4.10) of Ref.~\cite{Forseth:2015oua}. The angular momentum flux in Eq.~\eqref{eq:Ftailapproxresult} agrees with Eq. (7.2a) of Ref.~\cite{Arun:2009mc}. In these references, the tail terms are written in terms of ``eccentricity enhancement'' functions defined by
\bes
\label{eq:enhancementF} \bea
 \langle {\cal F}_{\rm tail}\rangle& =&\frac{128\pi}{5}\nu^2 x^{13/2}\varphi(e),\\
 \langle {\cal G}^z_{\rm tail}\rangle& =&\frac{128\pi}{5}\nu^2M x^{5}\tilde{\varphi}(e),
 \eea
 \ees
 where we identify from Eq.~(\ref{eq:averagedfluxes}) that
 \bes
 \label{eq:enhancement}
  \bea
 \varphi(e)&=& \frac{x^4J_{220}^2}{M^4}+\sum_{\substack{ k=1}}^\infty\sum_{m=-2}^2 \frac{15Ma_{2m}^2J_{2mk}^2}{256\pi\nu^2x^{13/2}}\, |\Omega_{mk}|^7\qquad\\
  \tilde \varphi(e)&=&\frac{x^4J_{220}^2}{M^4}+M^2\sum_{\substack{ k=1}}^\infty \left[\frac{J_{22k}^2\Omega_{2k}^6}{64 x^5}- \frac{\Omega_{-2k}^7J_{2-2k}^2}{64| \Omega_{-2k}| x^5}\,  \right].\qquad 
   \eea
 \ees
The explicit expressions for the terms in Eqs.~(\ref{eq:enhancement}) when specializing to the PN limit of the dynamics and  further restricting to small eccentricity can be read-off from Eqs.~\eqref{eq:Ftailapproxresult}. 
As an additional check on our results, we computed $\varphi(e)$ for high eccentricity by numerically evaluating the coefficients $J_{\ell m k}^{\rm Newt}$ and verified that we reproduce the results shown in Fig. 1 of Ref.~\cite{Arun:2007rg}.

\subsection{Gravitational waveform modes}
\label{sec:hlm}
The gravitational wave polarizations can be decomposed into spherical harmonic modes $h_{\ell m}$.  
To compute these modes we follow a similar procedure as for the fluxes. We use existing results for the instantaneous contributions, transform them to EOB coordinates, and perform the calculations for the hereditary terms from the multipolar expressions. For the leading order contribution, we verified explicitly that this procedure is in agreement with the results obtained by starting from the general multipolar post-Minkowski expansions, computing the radiative quadrupole moment in terms of $(e,p,\xi)$, differentiating twice with respect to time, and substituting the Newtonian EOB equations of motion to eliminate first derivatives. In future work, the direct use of the radiative multipole moments could also be examined as an alternative to using existing results for the $h_{\ell m}$ modes that have already been specialized to PN dynamics when eliminating higher than first order time derivatives. For the 1PN results given below, we used the instantaneous contributions to the modes that are provided in Ref.~\cite{Mishra:2015bqa} (see also Ref.~\cite{Gopakumar:2001dy} for previous work) as functions $h_{\ell m}(r, \phi, \dot r, \dot\phi)$ in modified harmonic coordinates. Since harmonic and ADM coordinates are equivalent at 1PN order  we employ the same transformation to EOB coordinates as for the fluxes that is given in Eqs.~\eqref{eq:deltasHtoEOB} and~\eqref{eq:phidotAtoE} and obtain the results as functions of $(e,p)$. These can be expressed in terms of $x=(M\omega_\phi)^{2/3}$ by working perturbatively to 1PN order and using Eq.~\eqref{eq:pofx}. The result for the $(2,2)$ mode is
\begin{widetext}
\bea
h_{22}^{\rm inst}&=&\frac{-8\sqrt{\pi}\ \mu}{\sqrt{5}D_L}e^{-2i\phi} \frac{x}{(1-e^2)}\Bigg[1+e\left(\frac{e^{-i\xi}}{4}+\frac{5e^{i\xi}}{4}\right)+\frac{e^2}{2} e^{2i\xi}+\frac{x}{(1-e^2)}\bigg\{- \frac{107}{42}+\frac{55\nu}{42} \nonumber\\
&&+e\left[\left(\frac{211}{168}\nu-\frac{383}{168}\right)e^{-i\xi}+\left(\frac{65}{24}\nu-\frac{121}{24}\right)e^{i\xi}\right]+ e^2 \left[\left(\frac{9 \nu }{28}-\frac{95}{168}\right) e^{-2 i
   \xi }+\left(\frac{52 \nu
   }{21}-\frac{673}{168}\right) e^{2 i \xi }+\frac{59
   \nu }{42}-\frac{115}{28}\right]\nonumber\\
   &&+e^3\left[\left(-\frac{13 \nu }{168}-\frac{199}{336}\right) e^{-i
   \xi }+\left(\frac{\nu }{28}+\frac{1}{112}\right)
   e^{-3 i \xi }+\left(\frac{13 \nu
   }{24}-\frac{143}{48}\right) e^{i \xi }+\left(\frac{5
   \nu }{4}-\frac{49}{48}\right) e^{3 i \xi }\right]\nonumber\\
   &&+e^4\left[\left(\frac{17 \nu }{84}-\frac{19}{28}\right) e^{2 i
   \xi }+\frac{1}{4} \nu  e^{4 i \xi }-\frac{\nu }{4}\right]\bigg\}\Bigg],\; \; \label{eq:h22Newt}
\eea
\end{widetext}
where $D_L$ is the luminosity distance of the source.
For circular orbits, this agrees with Eq. (79) in Ref.~\cite{Kidder:2007rt}. The other $(\ell,m)$ modes are computed in an analogous way, by inserting into the expressions from Ref.~\cite{Mishra:2015bqa} the transformations to EOB coordinates from Eqs.~\eqref{eq:deltasHtoEOB} and \eqref{eq:phidotAtoE} and re-expanding. We do not give the results explicitly here.

To compute the tail contributions we start from the relation between the $h^{\ell m}$ modes and the source's multipole moments 
\be
h^{\ell m}=\frac{1}{\sqrt{2}D_L}\bigg(U^{\ell m}-i V^{\ell m}\bigg), 
\ee
where
\be
U^{\ell m}=\frac{16\pi}{(2\ell+1)!!}\sqrt{\frac{(\ell+1)(\ell+2)}{2\ell(\ell-1)}}{\cal U}_L{\cal Y}^{\ell m^*}_L,
\ee
and $V^{\ell m}$ will not be needed at the order to which we are working here. The tail contributions to the radiative quadrupole at 1.5PN order are
\be
{\cal U}_{ij}^{\rm 1.5PN\, tail}=2M\int_0^\infty d\tau I_{ij}^{(4)}(t-\tau) \, {\rm \ln}\left(\frac{\tau}{b}\right).
\ee
Using Eqs.~(\ref{eq:IFourier}) and (\ref{eq:calYid}) this can be written as
\be
h_{2 m}^{\rm 1.5 PN}=c_m \sum_{k=-\infty}^\infty J_{\ell mk} \, \Omega_{mk}^4\, e^{-i(k\psi_r+m \psi_\phi)}\, {\cal I}(\Omega_{mk}),\ \ \ \ \label{eq:htail}
\ee
where $c_m=\sqrt{24}a_{2 m}M/D_L$ and ${\cal I}(\Omega_{mk})$ was given in Eq.~\eqref{calIdef}.

The result in Eq.~\eqref{eq:htail} is general but implicit. An explicit expression can be obtained when specializing to PN orbital dynamics and low eccentricity, similar to the case of the fluxes discussed above. The procedure was explained in detail in the previous section, and we refrain from repeating all the steps in detail here. To compare with existing PN results for circular orbits summarized e.g. in Ref.~\cite{Kidder:2007rt}, we express the phase in Eq.~\eqref{eq:htail} in terms of $\phi$ by using that $\psi_\phi=\phi-\Delta\phi_r$ as discussed in Sec.~\ref{sec:frequencies}. In the limit of  Newtonian orbital dynamics $\Delta \phi_r=\xi-\psi_r$ so that
\bea
h_{22}^{\rm1.5 PN}\mid_{\rm Newt}&=& \frac{\sqrt{24}a_{2 2}M}{D_L} e^{-2i\phi}\sum_{k=-\infty}^\infty J_{22k}\Omega_{2k}^4 \nonumber\\
&&\times e^{-2i\xi}e^{-i(k+2)\psi_r}{\cal I}(\Omega_{mk}). \label{eq:htail22test}
\eea
 The dependence on $\psi_r$ can be expanded in terms of $\xi$ by using Eq.~\eqref{eq:exppsirexp}. When writing out Eq.~\eqref{eq:htail22test} explicitly using the results derived in Sec.~\ref{eq:Newtexplicit} the expression becomes rather long. To give the results in a compact form for comparison to other work, we simplify the expressions by absorbing the imaginary contributions from  Eq.~(\ref{calIdef}) into a re-definition of the phase, as is described for circular orbits in Sec. IV C of Ref.~\cite{Kidder:2007rt}. To accomplish this, we first define a quantity $\ln(x_0)$ so that it eliminates all the dependences of $h_{\ell m}$ on $\gamma_E$ and the gauge parameter $b$ that come from substituting for ${\cal I}(\Omega_{mk})$ from Eq.~\eqref{calIdef}. The result is the same as for circular orbits given in Eq.~(68) of Ref.~\cite{Kidder:2007rt}: $\ln (x_0)=-2 [{\gamma_E}+{\ln( b)}]/3$.Using this in Eq.~(\ref{eq:htail}) with Eq.~(\ref{calIdef}) leads to a simpler expression for $h_{\ell m}^{\rm tail}$ that now involves $\ln (x/x_0)$ in place of $\ln(x)$. Next, we perform a shift in the phase at $O(x^{3/2})$ given by $\phi\to \tilde \phi=\phi+x^{3/2}\delta$ that is designed to eliminate all dependences on $\ln(x/x_0)$ when working perturbatively in the PN limit.   Performing this shift in the Newtonian term from Eq.~(\ref{eq:h22Newt}) and expanding for $x\ll1$ gives an imaginary contribution to the amplitude at $1.5$PN order parameterized by $\delta$. By choosing $\delta$ appropriately we can absorb the terms involving $\ln(x/x_0)$ in $h_{22}^{\rm tail}$ into $\tilde\phi$. Since the phase at Newtonian order scales as $O(x^{-5/2})$, the difference between $\phi$ and $\tilde \phi$ appears only at relative 4PN order and can be omitted in the approximation discussed here.  With all of these simplifications restricted to the limit of Newtonian orbital dynamics with low eccentricity, the result in Eq.~\eqref{eq:htail} becomes explicitly
\bea
&&h_{22}^{\rm tail}\approx -8\sqrt{\frac{\pi}{5}}\frac{\mu}{D_L}\,  x \, e^{-2i \phi}\; 2\pi x^{3/2}\nonumber\\
&&\qquad\times \bigg[1+e\left(\frac{11}{8}e^{-i\xi}+\frac{13}{8}e^{i\xi}\right)\nonumber\\
&&\qquad\quad+e^2\left(4+\frac{5}{8}e^{-2i\xi}+\frac{7}{8}e^{2i\xi}\right)+O(e^3)\bigg] ,\qquad \label{eq:htailsimple}
\eea
The circular-orbit limit of Eq.~\eqref{eq:htailsimple} reduces to the result given in Eq.~(79) of Ref.~\cite{Kidder:2007rt}. For building a refined EOB model in future work, this form~\eqref{eq:htailsimple} of the tail effects is in fact inadequate and instead, the tail contributions in their more general form from Eq.~\eqref{eq:htail} must be used. In addition, a re-summation of these terms will need to be performed in a similar way as for quasi-circular inspirals, where the leading order tails are re-summed into an exponential factor based on results from the test-particle limit. 

\subsection{Merger-Ringdown waveforms}
\label{sec:MR}
As is standard in the EOB formalism, see Ref.~\cite{ Bohe:2016gbl} for references on this,
a complete description of the waveform is obtained by connecting the inspiral-plunge signals described above to a merger-ringdown (RD) signal. To illustrate the method, we use the EOB prescription for quasi-circular binary coalescences developed in Ref.~\cite{ Bohe:2016gbl}, to which we refer the reader for all the details about the model. The merger-RD waveform uses simple analytic models for the amplitude $A_{22}$ and phase $\phi_{22}$ of the $({2,2})$ mode, with coefficients calibrated to NR data. A further input is the frequency of the least-damped quasinormal mode of the remnant, denoted by $\sigma_{220}$, whose computation involves a fitting formula for the mass and spin of the final object given the initial parameters that is also based only on circular inspirals. The merger-RD signal then takes the form
\be
h_{22}^{\rm merger-RD}=A_{22}(t-t_{\rm match})e^{i\phi_{22}(t-t_{\rm match})}e^{-i\sigma_{220}(t-t_{\rm match})},
\ee
where $t_{\rm match}$ is the time at which the inspiral and merger signals are matched. The complete signal then has the form
\be
h_{22}=h_{22}^{\rm insp-plunge}\theta(t_{\rm match}-t)+h_{22}^{\rm merger-RD} \theta(t-t_{\rm match}), 
\ee
where $h_{22}^{\rm insp-plunge}$ denotes the inspiral-plunge signal discussed in the previous subsections and $\theta$ is the Heaviside step function. 
For the circular EOB model, the matching time $t_{\rm match}$ occurs in the vicinity of the peak in the amplitude $|h_{22}|$, where information from NR that is included in the inspiral-plunge signal ensures the agreement with the NR values used to develop the merger-RD fit. Here, our inspiral model is limited to 1.5PN order and does not contain modifications coming from fitting the NR waveforms which are important around merger~\cite{Bohe:2016gbl}. We can resurrect those modifications 
for the systems that have circularized by the end of the inspiral. As for the binaries which still have a non-negligible 
eccentricity, similar corrections should be obtained from the analysis of the eccentric NR waveforms. 
Here we lack those important (around merger) features and we have decided to choose $t_{\rm match}$ by 
demanding smoothness in the instantaneous GW frequency. We start from the time where this frequency has maximum and 
move to earlier times searching for the instance which minimizes the jump in the first derivative of the GW frequency. 
This matching is sufficiently good as a proof of principle and has to be revised after comparison with NR data.

\section{Regularization near the turning points and adiabatic limit}
\label{sec:implementation}
In this Section we address considerations for the practical implementation of our model, and describe how to overcome  the numerical difficulties at the turning points of the radial motion. The numerical problem is that turning points give rise to terms of the form $0/0$. Analogous issues and explicit regularizations of these divergences are well-understood in the context of small mass-ratio binaries, see e.g. Ref.~\cite{Gair:2010iv} for details. Below, we discuss generalizations of these procedures that apply to the EOB dynamics. The idea is to derive an approximate form of the radial equation of motion that is manifestly finite at the turning points, and to switch between this representation and the exact expression. In general, a similar treatment is also necessary for the radiation reaction terms in the radial equation of motion, however, we explain below that for adiabatic inspirals, where only the orbit-averaged pieces of the fluxes drive the orbital evolution, these contributions vanish.  

\subsection{Regularizing the conservative dynamics}
We first consider the equation of motion for the angle $\xi$. It is convenient to consider the original expression from which it is derived through the re-parameterization: 
\bea
\dot{\xi} &=&{\cal P} - \frac1{\di r/\di \xi} \frac{\di r}{\di C_i} \dot{C}_i,
\label{E:rad_mot}\\
{\cal P}&=& \frac{\dot{r}}{\di r / \di\xi}. 
\eea
Here, the quantities $C_i$ are the set of constants for the conservative dynamics,  
either $\{P_{\phi}, E\}$ or $\{e, p\}$. The term ${\cal P}$ is given explicitly by 
\be
{\cal P} = \frac{\hat{P}_r }{ep\sin{\xi}} \frac{2A \mu M(1+e\cos{\xi})^2\left[r^2A D+2 M^2 Q_4\hat{P}_r^2\right]}{r^2E\left(E^2+2\nu M^2-M^2 \right)}.\label{E:xigeo}
\ee
 A numerical issue that arises at the turning points of the motion, where $\hat P_r=0$ and $\xi=0\, {\rm mod} \pi$, is that Eq.~\eqref{E:xigeo} becomes $0/0$. We now discuss a method to overcome this difficulty. We first note that in the expression for the radial momentum in Eq.~\eqref{eq:prof} and the definitions in Eq.~\eqref{eq:betadef}--~\eqref{eq:Ydef}, the function $\beta$ remains finite, however, the combination $Y$ goes to zero. This becomes apparent upon inserting the results of Eqs.~\eqref{eq:psofep}
to obtain
\bea
\label{eq:Yofr1r2}
Y&=&\frac{(1-e^2)^2}{M^2 p^2 r^2}\frac{1}{B}\bigg[\left(r_2^2-r_1^2\right)r^2A(r_1)A(r_2)\nonumber\\
&&\; \; \; +\left(r_1^2-r^2\right)r_2^2A(r_1)A(r) \nonumber\\
&&\; \; \; + \left(r^2-r_2^2\right)r_1^2A(r_2)A(r)\bigg],
\eea
where
\be
B=A\left[A_1(1-e)^2-A_2(1+e)^2\right], 
\ee
and we use the shorthand notation $A_{1,2}=A(r_{1,2})$.
From Eq.~\eqref{eq:Yofr1r2} it immediately follows that for $r\to r_{1,2}$ the function $Y$ vanishes. Near the turning points, the quantity $\hat P_r$ has the Taylor expansion
\be
\label{eq:prnearturning}
\hat{P}_r^2= \frac{Y}{DA} \left[ 1  - \frac1{4}\beta Y + \frac1{8} (\beta Y)^2  + O(\beta Y)^3\right].
\ee
To determine the general form of $Y$ near the turning points, we first re-express Eq.~\eqref{eq:Yofr1r2} in the form
\be
Y=\frac{p^2}{B}\left[A_1A_2(u_1^2-u_2^2)+AA_1(u^2-u_1^2)+A A_2(u_2^2-u^2)\right],
\ee
where $u=M/r$. 
Inserting the relations $u=(1+e\cos\xi)/p$ and $u_{1,2}=(1\mp e)/p$, and using trigonometric identities leads to
\bes
\be
\label{eq:Yexpression}
Y=B^{-1}\bigg[(A_2-A_1)A e^2\sin^2\xi+e \alpha\bigg],
\ee
with
\bea
\label{eq:alphadef}
\alpha
&=& 4 A_1\left[A-A_2\right] \cos^2\left(\frac{\xi}{2}\right)\nonumber\\
&&+4 A_2\left[A-A_1\right] \sin^2\left(\frac{\xi}{2}\right).
\eea
\ees
The first term in Eq.~\eqref{eq:Yexpression} already has a convenient form that will explicitly cancel the problematic divergence $\propto 1/\sin\xi$ in ${\cal P}$ in Eq.~\eqref{E:xigeo} upon using the expansion~\eqref{eq:prnearturning}. 
To proceed further with manipulating the function $\alpha$ to obtain a manifestly finite expression for $Y$ near the turning points requires specializing to a particular form of the potentials. 
In general, the EOB A-potential in either the Taylor expanded or the log-resummed, calibrated version involves polynomial and logarithmic functions of $u$ and has the general form
\be
A=\sum_{k=0}^{k_{\rm max}} a_k u^k+\left[b+g\,u^s \right] \log\left[f(u)\right], \label{eq:Adecomp}
\ee
where $(a_k, b, g)$ are constants. The values of all the coefficients and functions in Eq.~\eqref{eq:Adecomp} for particular choices of the potential are given in the appendix. The difference $A-A_i$ can then be written as
\bea
A-A_i&=&(u-u_i)\sum_{k=0}^{k_{\rm max}}\sum_{\ell=0}^{k-1}a_k u^\ell u_i^{k-\ell-1}\nonumber\\
&&+\frac{g}{2}(u^s-u_i^s)\log\left(f\, f_i\right)\nonumber\\
&&+\left[b+\frac{g}{2}\left(u^s+u_i^s\right)\right]\log\left(\frac{f}{f_i}\right), \label{eq:Adiffexpand}
\eea
where we have used that
\bea
&&u\log{f(u)} - u_i\log{f(u_i)} =  \\ 
&&\quad\frac1{2}\left[ 
(u-u_i)\log{(f(u)f(u_i))} + (u+u_i)\log{\frac{f(u)}{f(u_i)}} 
\right]. \nonumber
\eea
Since 
\be
u -u_1=\frac{2e}{p}\cos^2\left(\frac{\xi}{2}\right), \ \ \ \ u -u_2=-\frac{2e}{p}\sin^2\left(\frac{\xi}{2}\right),
\ee
and $4\cos^2\left(\xi/2\right)\sin^2\left(\xi/2\right)=\sin^2\xi$, we see that the polynomial terms in Eq.~\eqref{eq:Adiffexpand} when used in Eq.~\eqref{eq:alphadef} will straightforwardly combine into the desired form $\propto \sin^2\xi$ without further manipulations. For the first logarithmic term, in the second line of Eq.~\eqref{eq:Adiffexpand}, the prefactor $\propto (u^s-u_i^s)$ can be factored into $(u-u_i)\sum (\ldots )$ similar to the polynomial terms in the first line of Eq.~\eqref{eq:Adiffexpand}. Finally, the terms in the last line of Eq.~\eqref{eq:Adiffexpand} do not have an explicit decomposition that would combine into $\sin^2\xi$ in Eq.~\eqref{eq:alphadef}. However, the functions $f(u)$ relevant here consist only of powers and logarithms of $u$ (see Appendix). Therefore, we can employ a series expansion for $\log[f(u)/f(u_i)]=\log(1+\Delta_i)$, where $\Delta_i=|f(u)-f(u_i)|/[f(u_i)^(i-1)f(u)^{2-i}]\leq 1$.  As shown in the Appendix, $\Delta_i\propto (u-u_i)$, which provides the desired factor to cancel the divergence. Performing all of these manipulations leads to the following result for $Y$:
\be
\label{eq:Yapprox}
Y =\frac{2e^2\sin^2{\xi}}{p} \frac{A_2 dA_1(u)-A_1 dA_2(u)  + pA(u) (A_2-A_1)/2}{A(u)\left[ A_1(1-e)^2 - A_2(1+e)^2\right]}. \qquad 
\ee
The functions $dA_i(u)$ depend on the form of the potential (re-summed or Taylor expanded) and are given in the Appendix. 

Finally, substituting Eq.~\eqref{eq:Yapprox} into expression 
\eqref{eq:prnearturning} shows the explicit cancellation of $\sin{\xi}$ in \eqref{E:xigeo} for the motion close to the turning points. 

In a numerical code one can then switch between using the full expression for $\hat P_r$ from Eq.~\ref{eq:prof} to compute ${\cal P}$ in \eqref{E:xigeo}, and employing the manifestly finite expansion from Eqs.~\eqref{eq:prnearturning} and \eqref{eq:Yapprox} when the motion approaches one of the turning points. 

A similar regularization as discussed above is also necessary for the radiation reaction contributions to $\dot \xi$. However, as we will explain below, these contributions vanish for adiabatic inspirals which are the main focus of our implementation here. Hence, we defer the details of the regularization for these instantaneous radiation reaction terms to Appendix~\ref{sec:rrterms}.

\subsection{Adiabatic limit}
\label{sec:forces}
In this paper we limit the computation of explicit results to adiabatic waveforms. By ``adiabatic" we mean that only the averaged radiation reaction forces instead of the full instantaneous forces are used to evolve the trajectory. This approximation has the following consequences. The energy and angular momentum balance relations, as discussed in detail in Ref.~\cite{Bini:2012ji}, are given by
\be
\dot E=-{\cal F}-\dot E_{\rm Schott}, \ \ \ \ \ \dot P_\phi=-{\cal G}^z-\dot L_{\rm Schott}. \label{eq:balance}
\ee
Here, the Schott energy and angular momentum $\{E,L\}_{\rm Schott}$ represent the
interaction of the system with the near-zone field. The fluxes ${\cal F}$ and ${\cal G}^z$ are the combined fluxes at infinity and through the horizon for black holes, however, in this paper, as in standard PN computations, we consider only the fluxes at infinity. As discussed in~\cite{Bini:2012ji} one can always choose the gauge freedom that $\dot{L}_{\rm Schott} = 0$.

Since the Schott terms have only been derived to 2PN order for the instantaneous terms~\cite{Bini:2012ji} and the 1.5PN hereditary contributions remain unknown, we will specialize here to adiabatic radiation reaction forces that are expected to be a reasonably good approximation in the regime when the orbital timescales are short compared to the radiation reaction timescale in the sense that they describe the dominant effects. The  change in $E$ over a finite time interval due to $\dot E_{\rm Schott}$ is oscillatory and averages to zero over a generic orbit; its effect on the dynamics is therefore smaller than other, secularly growing pieces in Eq.~\eqref{eq:balance}. This argument breaks down non-adiabatic  regions which, in generic dynamical systems, generally comprise either resonances or separatrices~\cite{1982nods.book.....G,1988macc.book.....A}. Since in this paper, we consider only nonspinning binaries, possible resonances could occur between $\omega_r$ and $\omega_\phi$. However, the rotational motion in $\phi$ has a different status than the radial librations characterized by $\omega_r$, see e.g. Refs.~\cite{1978mmcm.book.....A, Schmidt:2002qk,Hinderer:2008dm}. Consequently, resonances between $\omega_r$ and $\omega_\phi$ do not lead to the large corrections to the fluxes that occur for resonances between two librational frequencies, where instantaneous terms that are normally oscillatory become stationary and thus behave like orbit-averaged contributions~\cite{Flanagan:2010cd}. The absence of sudden large corrections at resonance can also be seen from the time-dependence of the instantaneous fluxes, given e.g. for the tail terms in Eq.~\eqref{eq:FGtailinst}, which involve only $\psi_r=\omega_rt$ and are independent of $\psi_\phi$. For the nonspinning binaries considered here, the only expected non-adiabatic regions are therefore the two separatrices in the phase space of the radial motion where the behavior of $\psi_r$ changes: highly eccentric systems close to unbound orbits, and the end of the inspiral close to the plunge. For the purpose of this paper we will specialize to adiabatic contexts with a simplified treatment of the transition to the plunge, and leave more detailed studies of the instantaneous forces and the validity of adiabatic waveforms to future work. For estimates of the effect of the oscillatory terms on the GW phase we refer the reader e.g. to Ref.~\cite{Moore:2016qxz}, where this issue is considered in the PN context and low-eccentricity limit in Sec.V and Fig. 3 therein.  

In the adiabatic limit, the evolution is driven by the orbit-averaged radiation reaction forces so that
\be
\label{eq:adiabaticEdot}
 \dot E^{\rm adiab}=-\langle{\cal F}\rangle, \ \ \ \ \ \dot P_\phi^{\rm adiab}=-\langle{\cal G}^z\rangle.
\ee

A further consequence of the adiabatic approximation is that the radiation reaction contributions to Eq.~\eqref{eq:xidotfinal} vanish. This can be verified either by explicit computations of the orbit-average of the radiation reaction terms in the equation of motion for $\xi$, Eq.~\eqref{eq:xidotgen}, or by using similar theoretical arguments about the properties of the dissipative piece of the forcing functions on the phase variables in the test-particle limit~\cite{Hinderer:2008dm,Gair:2010iv}. This implies that
\bes
\label{eq:adiabaticeom}
\be
\label{eq:xidotadiab}
\dot \xi^{\rm adiab}={\cal P}(\xi, e,p),
\ee
where the parameters $e$ and $p$ are time-dependent and their evolution is given by
\bea
\dot{e}^{\rm adiab}&=&-\frac{c_{Ep}}{\mu}\langle {\cal G}^z\rangle+c_{Lp}\langle {\cal F}\rangle,\\
\dot{p}^{\rm adiab}&=&\frac{c_{Ee}}{\mu}\langle {\cal G}^z\rangle-c_{Le}\langle {\cal F}\rangle,
\eea
\ees
with the coefficients given in Eq.~\eqref{eq:cAbdef} and computed using Eq.~\eqref{eq:psofep}.

With regards to future refinements of the eccentric EOB model, the higher order PN corrections to $\langle {\cal F}\rangle$ and $\langle {\cal G}^z\rangle$ can readily be obtained in the following way. One starts from the existing PN results for these quantities that are usually given in terms of $(x,e_t)$ and makes use of the known relation to the gauge invariant quantities $e_t(\epsilon,j)$. In these relations one substitutes the PN expansion of $(\epsilon,j)$ and of $x$ in terms of the EOB variables $(e,p)$, which is straightforward to compute by expanding the relativistic EOB results, and re-expands the results. For the hereditary terms, in the approximation of PN orbital dynamics but for arbitrarily high eccentricity, a similar mapping from $(e_t,x)$ to the EOB variables can be applied to the eccentricity re-summed results of the hereditary terms that were calculated in Ref.~\cite{Loutrel:2016cdw}; the alternative re-summations of Ref.~\cite{Forseth:2015oua} could also be directly used. However, nontrivial computations based on the methods developed in this paper will be necessary to obtain the $h_{\ell m}$ modes because they also involve harmonics of the phase variables that are distinct between PN and EOB. These results for the modes then have to be re-summed using a similar strategy as has been employed for quasi-circular EOB waveforms , see e.g. Ref.~\cite{Damour:2008gu}. The most convenient way to express the results would be in terms of eccentricity enhancement factors to the circular-orbit limit of the $h_{\ell m}$-modes that would involve not only the eccentricity itself but also harmonics of the radial phase. Once the EOB $h_{\ell m}$ modes have been constructed, a self-consistent EOB model will employ these modes to obtain the fluxes, which is another important aspect of future work that we did not address in this paper.

For binaries whose orbits retain a substantial eccentricity at the end of the inspiral a further treatment is necessary. The transformation to $(\dot{e},\dot{p})$ from Eq.~\eqref{eq:edotpdot} becomes singular at the transition to the plunge. This issue is known from evolutions of orbits in Schwarzschild ~\cite{Cutler:1994pb} and Kerr spacetimes~\cite{Sundararajan:2008bw}, and can likewise be treated by a local analysis of the behavior in this regime. A refined treatment will be the subject of future work. Finally, we note that the limit of circular orbits, $e\to 0$, a further regularization is necessary which will also be addressed in future work.

\section{Results and Discussion}
\label{sec:results}
In this section we first summarize our current proof-of-principle implementation of the model present a few illustrative results, all based on the foundations for the EOB model that we have developed in this paper. The purpose is to demonstrate the practical use of our method, and to exhibit features of the relativistic parameterization when compared to 1PN expansions. More detailed and comprehensive studies, and comparisons with NR, full PN and other eccentric waveform models will be the subject of future work. 

\subsection{Summary of equations used}
\subsubsection{EOB inspiral trajectory}
In the current implementation to demonstrate the practical use of the methods, we first compute EOB inspirals by using Eqs.~\eqref{eq:adiabaticeom} together with Eqs.~\eqref{eq:fullset} and the regularization discussed in Sec.~\ref{sec:implementation} and then calculate the $h_{\ell m}$ modes for that trajectory. Specifically, we solve the system
\bes
\bea
\dot \xi&=& \Bigg. \frac{ A \hat{P}_r(1+e \cos \xi )^2\sqrt{(1+e)^2 A(r_2)-(1-e)^2 A(r_1)}}{2 e^{3/2}p^3 \sin\xi \, E \sqrt{A(r_1)A(r_2)}}\nonumber\\
&&  \times \left[p^2AD+2\hat{P}_r^2(1+e \cos \xi )^2  Q_4\right], \Bigg.\ \ \ \ \ \ \ \ \ \\
 \dot\phi&=&\Bigg. \frac{A \sqrt{A(r_1)-A(r_2)} (1+e \cos \xi)^2}{2 \sqrt{e} \, p \, E \sqrt{A(r_1)}
   \sqrt{A(r_2)} },\\
\dot{e}&=&-\frac{c_{Ep}}{\mu}\langle {\cal G}^z\rangle+c_{Lp}\langle {\cal F}\rangle,\\
\dot{p}&=&\frac{c_{Ee}}{\mu}\langle {\cal G}^z\rangle-c_{Le}\langle {\cal F}\rangle,
\eea
with $r_{1,2}=pM/(1\mp e)$, $Q_4=2\nu(4-3\nu)$, and
\bea
\hat{P}_r^2&=&\frac{M^2p^2 A D}{2(1+e\cos\xi)^2Q_4}\bigg[-1 \nonumber \\
&& +\sqrt{1+\frac{4(1+e\cos\xi)^2Q_4}{M^2 p^2 A D} Y}\,\bigg],\\
Y&=&\frac{\left(E^2+2\nu M^2-M^2\right)^2}{4\mu^2 M^2 A}-1-\frac{\hat{P}_\phi^2}{r^2}. \qquad\\
\frac{ E^2}{M^2}&=&1-2\nu+\frac{4\nu \sqrt{e}\sqrt{A(r_1)A(r_2)}}{\sqrt{(1+e)^2A(r_2)-(1-e)^2A(r_1)}},\qquad \\
\hat{P}_\phi^2&=&\Bigg. \frac{p^2 M^2 \left(A(r_2)-A(r_1)\right)}{(1-e)^2A(r_1)-(1+e)^2A(r_2)}\Bigg. .
\eea
 For the EOB potentials $A$ and $D$ we use the log-resummed expressions given in Appendix~\ref{sec:logpot}. The coefficients $c_{Ab}$ are computed by differentiating the expressions for $(E,\hat{P}_\phi)$ according to
\be
c_{Cb}=\frac{\partial C/\partial b}{(\partial E/\partial p)(\partial \hat{P}_\phi/\partial e)-(\partial E/\partial e)(\partial \hat{P}_\phi/\partial p)},
\ee
\ees
where $C=\{E,\hat{P}_\phi\}$ and $b=\{e,p\}$. 
For the fluxes, we use for the illustrations in this section the explicit but only approximate expressions computed in Sec.~\ref{sec:fluxes} that are given by
\bea 
\langle{\cal F}\rangle &=&\frac{32\mu^2(1-e^2)^{3/2}}{5 p^5M^2}\bigg\{1+\frac{73}{24}e^2+\frac{37}{96}e^4\\
&&+\frac{1}{p}\bigg[-\frac{1247}{336}-\frac{5\nu}{4}-e^2\left(\frac{9181}{672}+\frac{325\nu}{24}\right)\nonumber\\
&&+e^4\left(\frac{809}{128}-\frac{435\nu}{32}\right)+e^6\left(\frac{8609}{5376}-\frac{185\nu}{192}\right)\bigg]\bigg\}\nonumber\\
&&+\frac{128\nu^2 \pi(1-e^2)^{13/2}}{5p^{13/2}} \left[1+\frac{2335 e^2}{192}+\frac{42955 e^4}{768}\right],\nonumber\\
\langle {\cal G}^z\rangle&=&\frac{32\left(1-e^2\right)^{3/2}\mu^2}{5M p^{7/2}}\bigg\{1+\frac{7}{8}e^2\nonumber\\
&&+\frac{1}{p}\left[-\frac{1247}{336}-\frac{7\nu}{4}-e^2\left(\frac{425}{336}+\frac{401\nu}{48}\right)\right.\nonumber\\
&&\left.+e^4\left(\frac{10751}{2688}-\frac{205\nu}{96}\right)\right]\bigg\}\nonumber\\
&&+\frac{128M\nu^2 \pi (1-e^2)^{5}}{5p^5}\left[1+\frac{209 e^2}{32}+\frac{2415 e^4}{128}\right],
\eea

\subsubsection{Regularization, circular-orbit limit, and plunge}

The regularization for the numerical issues near the turning points was explained in Sec.~\ref{sec:implementation}. We use the approximate form of ${\cal P}$ (the right hand side of $\dot \xi$) given by substituting Eq.~\eqref{eq:Yapprox} for $Y$ with the explicit relations given in Appendix~\ref{sec:logpot} into the expression 
\eqref{eq:prnearturning} for $\hat P_r$ when $\xi$ is within $10^{-2}$ of the turning points $(0,\pi)\, {\rm mod}\, 2\pi$. We also implemented the equations of motion in Mathematica and found that in that case the regularization is not necessary to obtain solutions, and the results agree with those produced with our python code. 

To avoid divergences in the circular-orbit limit we switch to the equations of motion for $(r,\phi, E, \hat{P}_\phi)$ given in Eqs.~\eqref{eomsEOBrphiEL}. We arbitrarily choose to perform this change in the description for $e\lesssim 5\times 10^{-3}$; a thorough treatment of the limit of vanishing eccentricity will be developed in future work. For the radiation reaction forces in Eqs.~\eqref{eomsEOBrphiEL} we use $F_E=-\langle {\cal F}\rangle$ and $\hat{ F}_\phi=-\langle {\cal G}^z\rangle/\mu$ with the fluxes specialized to circular orbits. Similarly, we also specialize the $h_{\ell m}$ modes to circular orbits in this case. 

Care is also required in the cases where the motion reaches the eccentric separatrix between inspiral and plunge. To describe the evolution in this regime we follow the treatment discussed in the context of extreme mass ratio binaries in Ref.~\cite{Sundararajan:2008bw}. This is based on expressing the radial equation from~\eqref{eomsEOBrphiEL} in terms of $(r, \dot r, E, \hat P_\phi)$ as 
\be
\label{eq:Vrdef}
\dot{r}^2 \equiv V_r (r, C_i), \ \ \ \ \ \ V_r=X^2 \hat P_r^2,
\ee
The function $\hat{P}_r^2(r,C_i)$ is given in \eqref{eq:prof}--\eqref{eq:Ydef} and the quantity $X$ can be read off directly from Eq.~\eqref{eq:rdoteom}:
\be
\label{eq:Xexpr}
X= \frac{2A \mu M^2\left[r^2A D+2 M^2 Q_4\hat{P}_r^2\right]}{r^2E\left(E^2+2\nu M^2-M^2 \right)}.
\ee
The separatrix is the solution to $V_r=dV_r/dr=d^2 V_r/dr^2=0$ and is the innermost stable orbit for bound motion. To define the near-separatrix region we use a similar criterion as described in Ref.~\cite{Sundararajan:2008bw}. We also follow the prescription for the evolution in this regime from Ref.~\cite{Sundararajan:2008bw} and do not repeat the details here. The idea of the method is to take a time derivative of Eq.~\eqref{eq:Vrdef} and perform a Taylor expansion around the separatrix values. Similarly, the evolution of $(E, \hat P_\phi)$ is obtained from a Taylor expansion of the fluxes around their values the separatrix. This is a highly non-linear and dynamic 
regime of binary evolution and requires a more careful treatment using inputs from NR data, which we delegate to the future work.

\subsubsection{Waveforms}
Given the EOB inspiral trajectory we calculate the waveform modes as explain in Sec.~\ref{sec:hlm}. For simplicity we consider only the $(2,2)$ mode here. For the inspiral part we use Eqs.~\eqref{eq:h22Newt} and the simplified approximate result in Eq.~\eqref{eq:htailsimple}, with refinements deferred to future work. We compute $x$ by numerically integrating Eq.~\eqref{eq:omegaphidef} for the EOB trajectory. 
For the plunge part of the GW signal we use the expressions for the instantaneous contributions to $h_{22}$ in terms of $(r, \dot r, \phi, \dot\phi)$, together with the circular-orbit limit of the tail contributions. The merger-ringdown signals and their attachment procedure were explained in Sec.~\ref{sec:MR}.

\subsection{Illustrative results}

\begin{figure}
\includegraphics[width=\columnwidth]{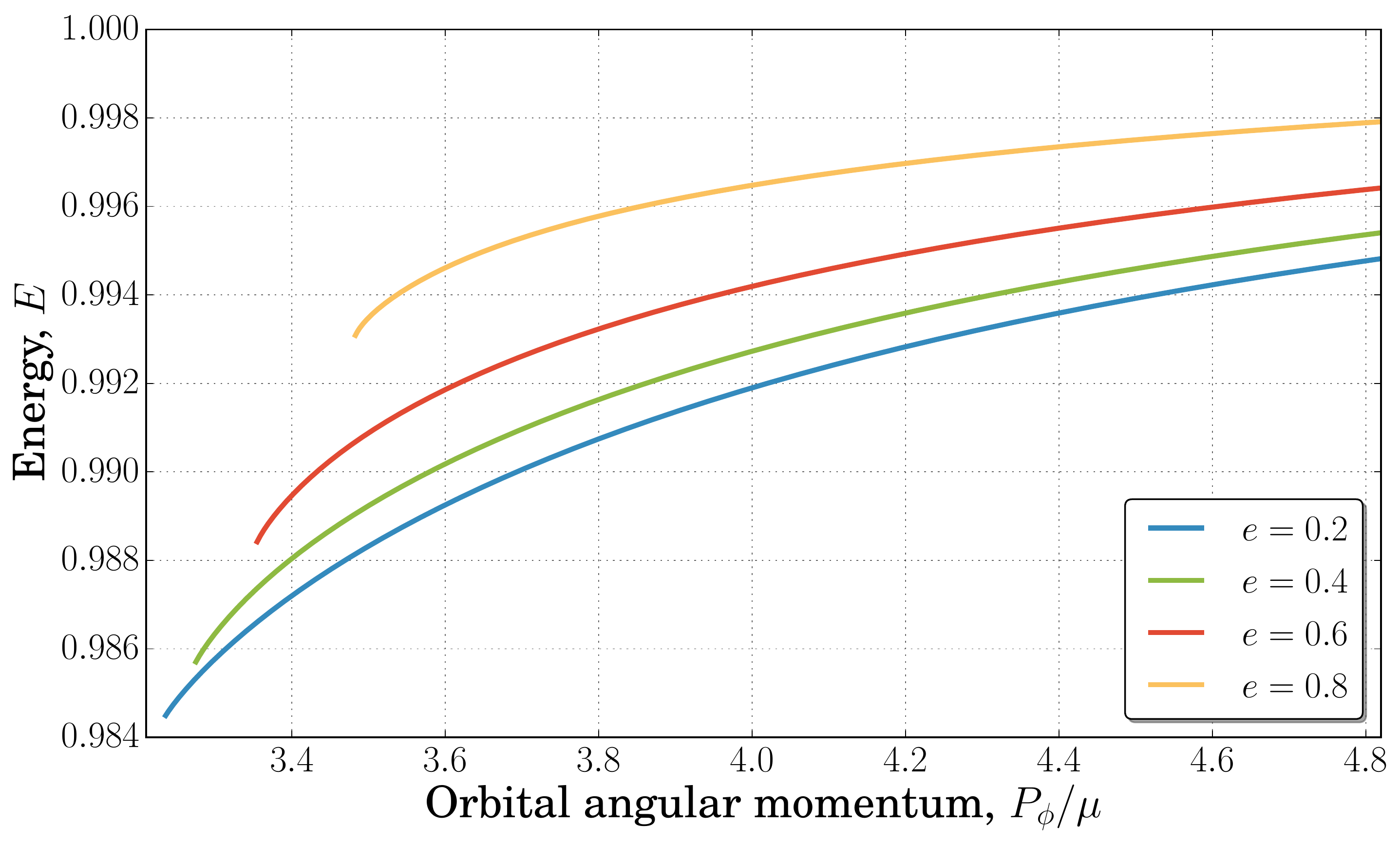}
\caption{\emph{Impact of eccentricity on the energy versus angular momentum of a binary}. Shown are the results for the energy as a function of the orbital angular momentum for mass ratio 2 and computed from the conservative dynamics for eccentricities of $e=0.8$ (orange-yellow curve), $e=0.6$ (red curve), $e=0.4$ (green curve), and $e=0.2$ (blue curve). The curves terminate at the innermost stable orbit. This illustrates that for a given angular momentum orbits with high eccentricity have a higher energy than those with lower eccentricity. }
\label{fig:EJvaryecc}
\end{figure}

\begin{figure}
\includegraphics[width=\columnwidth]{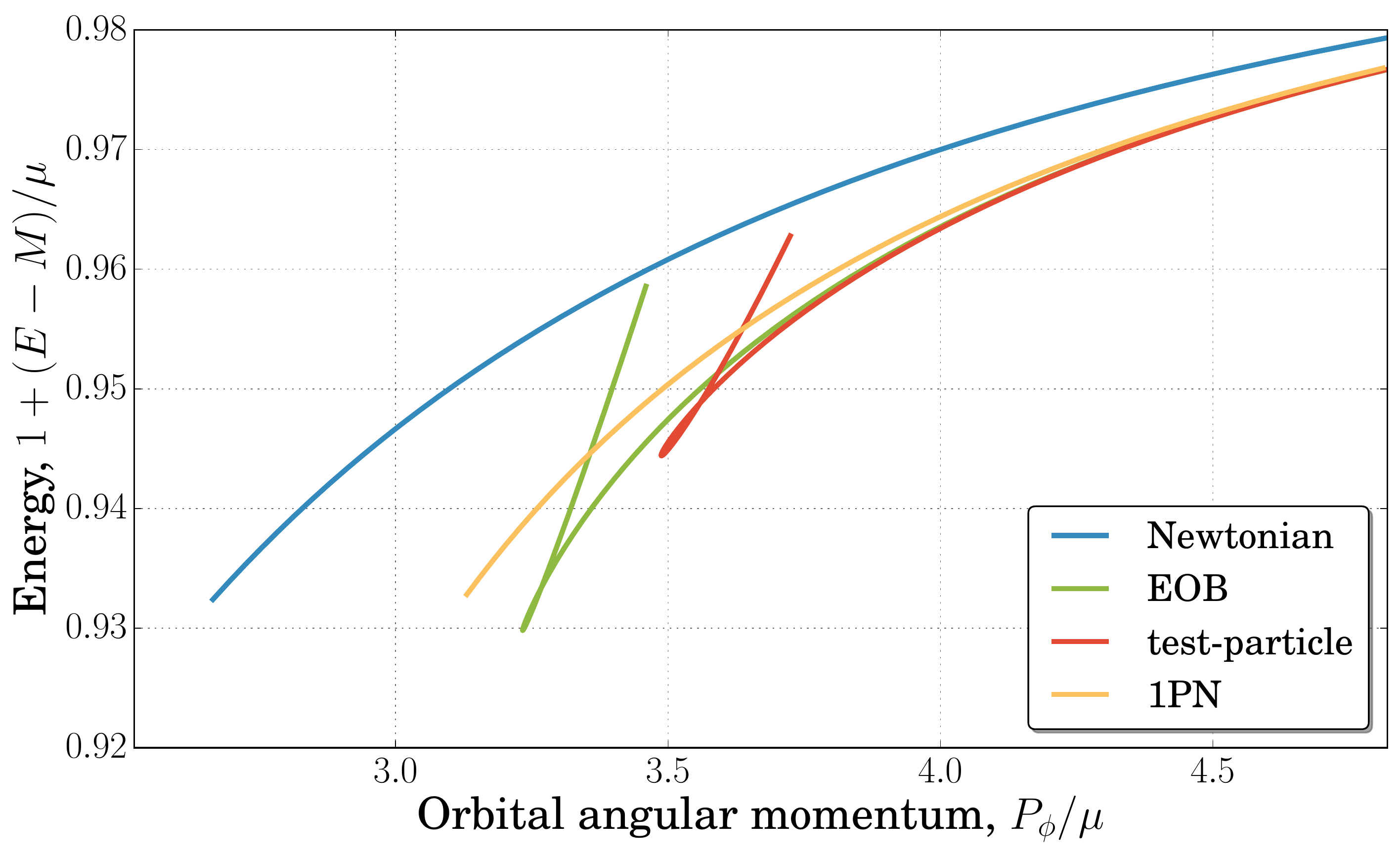}
\caption{\emph{Energetics of the conservative EOB dynamics in different approximations}. The plot illustrates the results for the rescaled energy versus reduced orbital angular momentum for a system with mass ratio 2 and eccentricity of $e=0.2$. The different curves correspond to the Newtonian limit (blue line), the 1PN expansion of the EOB results (orange-yellow curve),
the test-particle limit (red), and the EOB model (green). The cusps in the EOB and test-particle limit curves occur at the last stable orbit; this feature is absent from the Newtonian and 1PN curves.} 
\label{fig:EJEOBvsPNTPL}
\end{figure}

The first aspect we consider are the energetics of the binary in the absence of radiation reaction. The EOB model for the conservative dynamics contains no approximations in the eccentricity and is fully relativistic. Figure~\ref{fig:EJvaryecc} illustrates the effect of eccentricity on the energy plotted as a function of the orbital angular momentum per unit reduced mass. At a fixed angular momentum orbits with higher eccentricity have a higher energy than those with lower eccentricity. An energy of $E=1$ would correspond to a marginally bound parabolic orbit. The curves terminate at the innermost stable orbit (iso) which is the eccentric separatrix discussed above. A comparison between the EOB energetics and those for a PN or test-particle orbit is shown in Fig.~\ref{fig:EJEOBvsPNTPL} for a fiducial system with $e=0.2$ and mass ratio $2$. This illustrates that in the strong-field regime, corresponding to low angular momentum in the plot, the EOB results differ from both the 1PN and test-particle limit results. 

\begin{figure}
\includegraphics[width=\columnwidth]{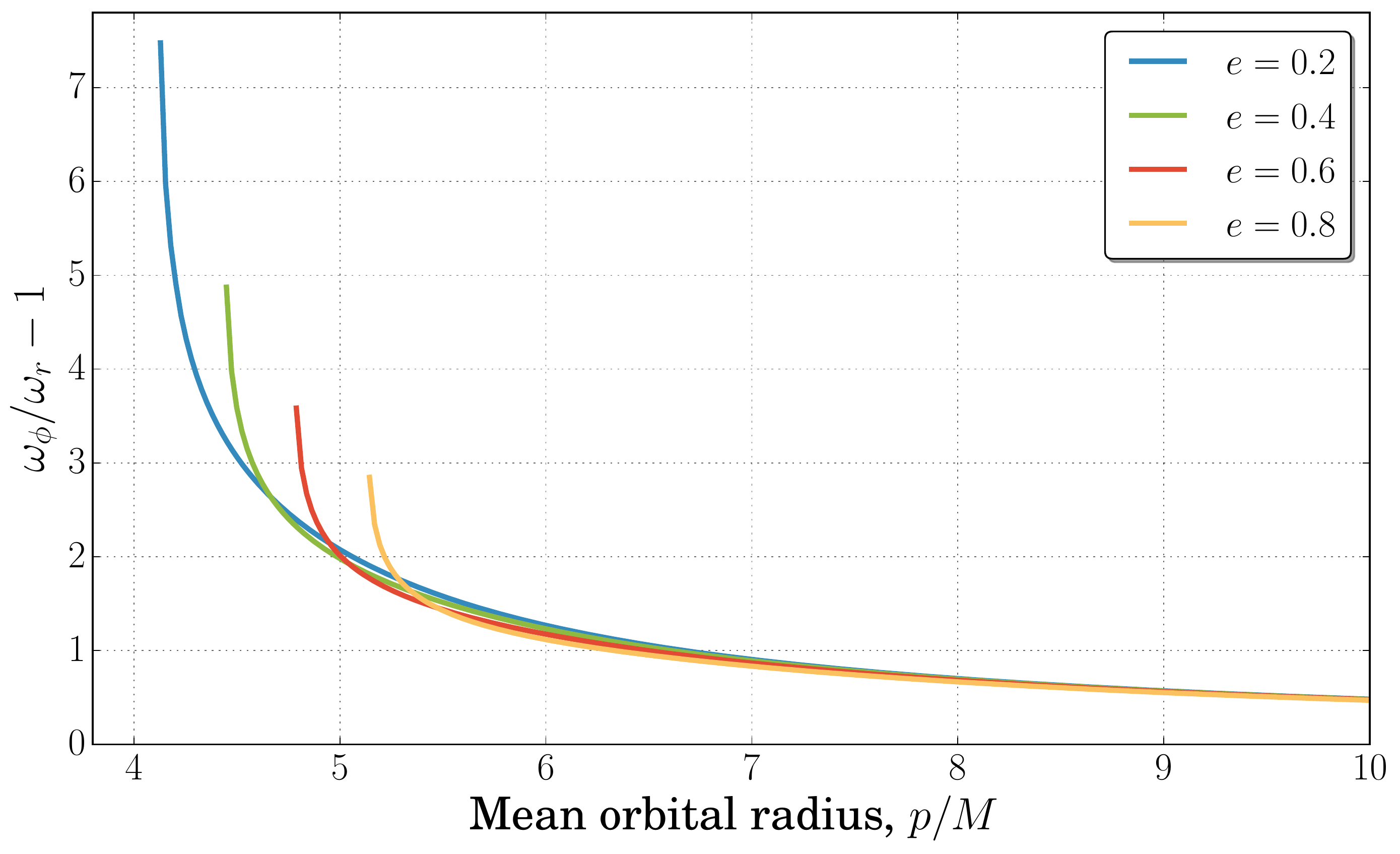}
\caption{\emph{Effect of eccentricity on periastron precession}. The results shown are for an equal-mass binary, and the smallest $p$ for each curve is $p_{\rm iso}+0.01$ since the radial frequency approaches zero close to the iso. As the eccentricity increases, the shape of the orbit becomes more elliptical and the orbits spend more time in the weak-field regime. For orbits of the same mean orbital radius those with higher eccentricity therefore have a smaller periastron precession. }
\label{fig:frequenciesofp}
\end{figure}

Next, we consider the features of the radial and azimuthal frequencies computed from the conservative dynamics.  The effect of eccentricity on these frequencies as functions of the mean orbital separation is shown in Fig.~\ref{fig:frequenciesofp}. As the eccentricity increases, orbits become more elliptical and on average spend more time in the weak-field region than nearly circular orbits with the same mean radius. This leads to a decrease of the quantity characterizing the periastron precession $k=-1+\omega_\phi/\omega_r$ since in the Newtonian limit $\omega_r=\omega_\phi$.   In contrast to the QK formalism, the parameterization employed here has no restrictions on the size of the periastron precession and therefore applies even for zoom-whirl orbits that occur in the vicinity of the separatrix.  

\begin{figure}
\includegraphics[width=\columnwidth]{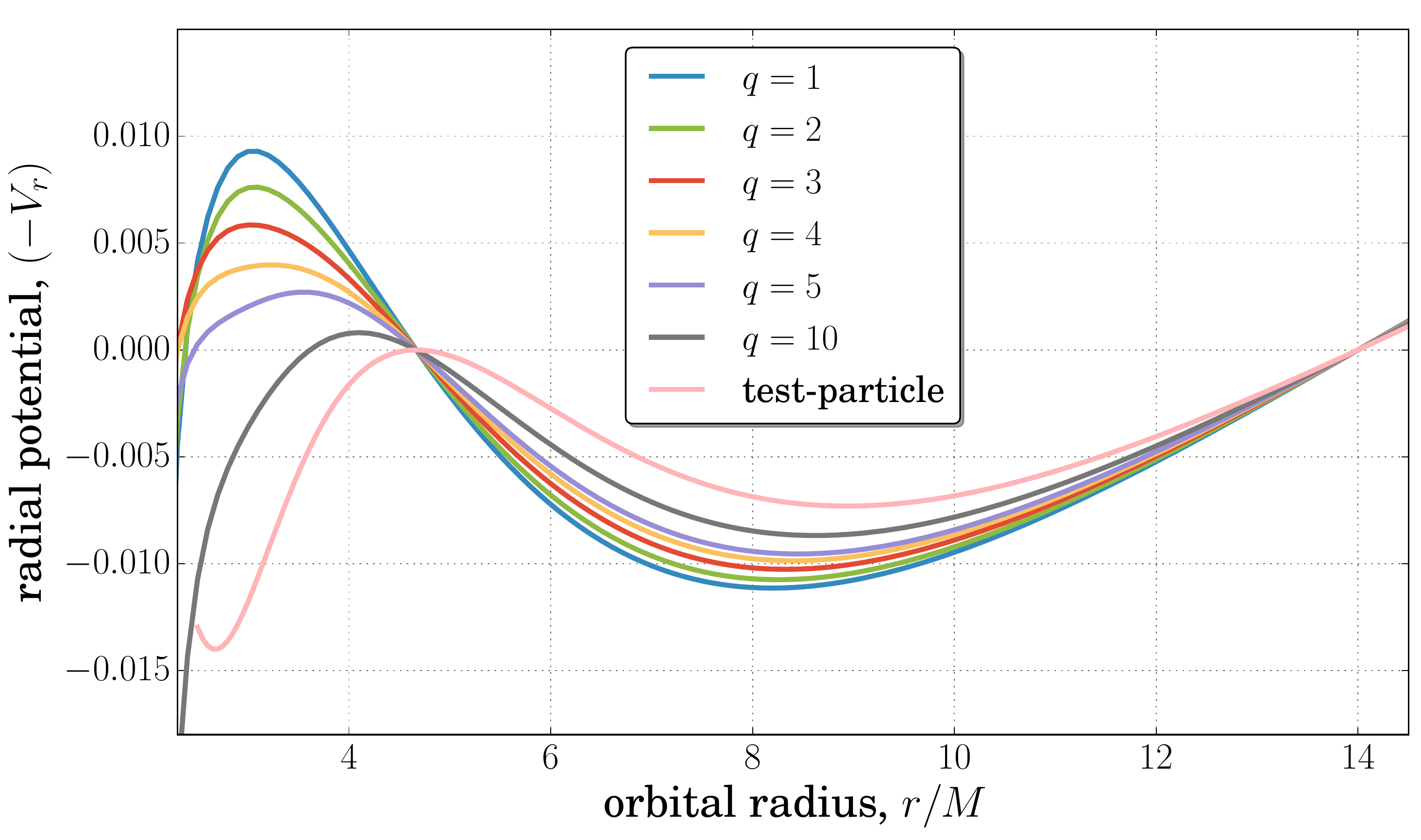}
\caption{\emph{Variation of the shape of the EOB effective radial potential with the mass ratio.} Turning points of bound orbits correspond to roots of the potential $V_r$, shown here for orbits with $e=0.5$ and $p=7$. We observe that a maximum in $(-V_r)$ corresponding to a homoclinic orbit in the radial motion is present for all mass ratios. }
\label{fig:potential}
\end{figure}

The EOB model employed here exhibits similar features to those observed for the evolution of test-particle orbits in Schwarzschild spacetime. One such characteristic is the presence of a separatrix demarcating the inspiral and plunge that is analogous to the curve $p=6+2 e$ for geodesics and corresponds to orbits whose periapsis is at the maximum of the effective radial potential $V_r$ defined by Eq.~\eqref{eq:Vrdef}.
 The features of the radial potential for different mass ratios are illustrated in Fig.~\ref{fig:potential}, for a fiducial choice of $(e,p)=(0.5, 7)$. 

\begin{figure}
\includegraphics[width=\columnwidth]{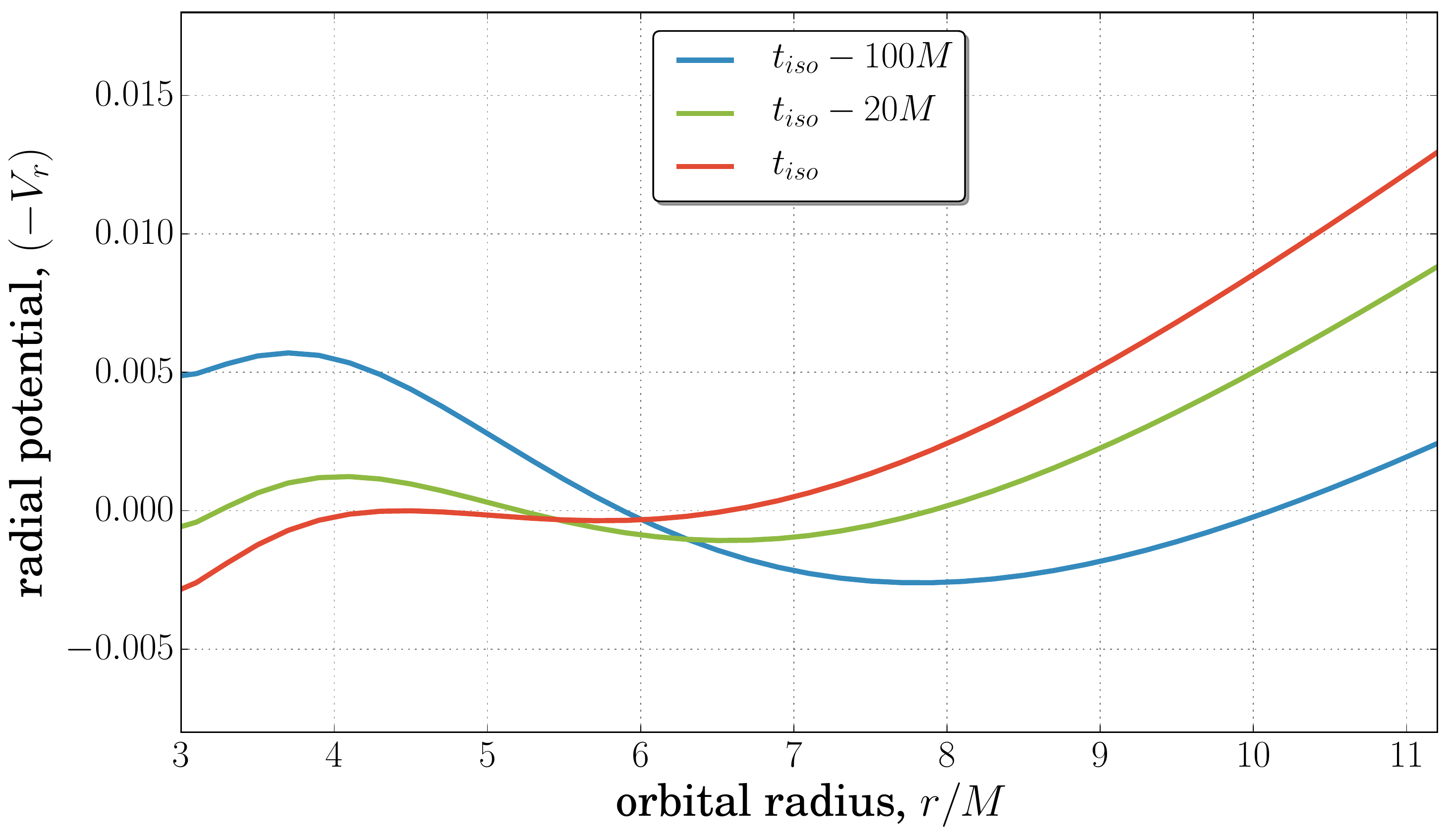}
\caption{\emph{Evolution of the EOB radial potential $(-V_r)$ during an inspiral. }The times of the snapshots are the instant when trajectory crosses the separatrix or innermost stable orbit (iso) of the corresponding conservative dynamics (red curve), and $20$M (green) and $100$M (blue curve) prior to this time.  }
\label{fig:potentialsnapshot}
\end{figure}

\begin{figure}
\includegraphics[width=\columnwidth]{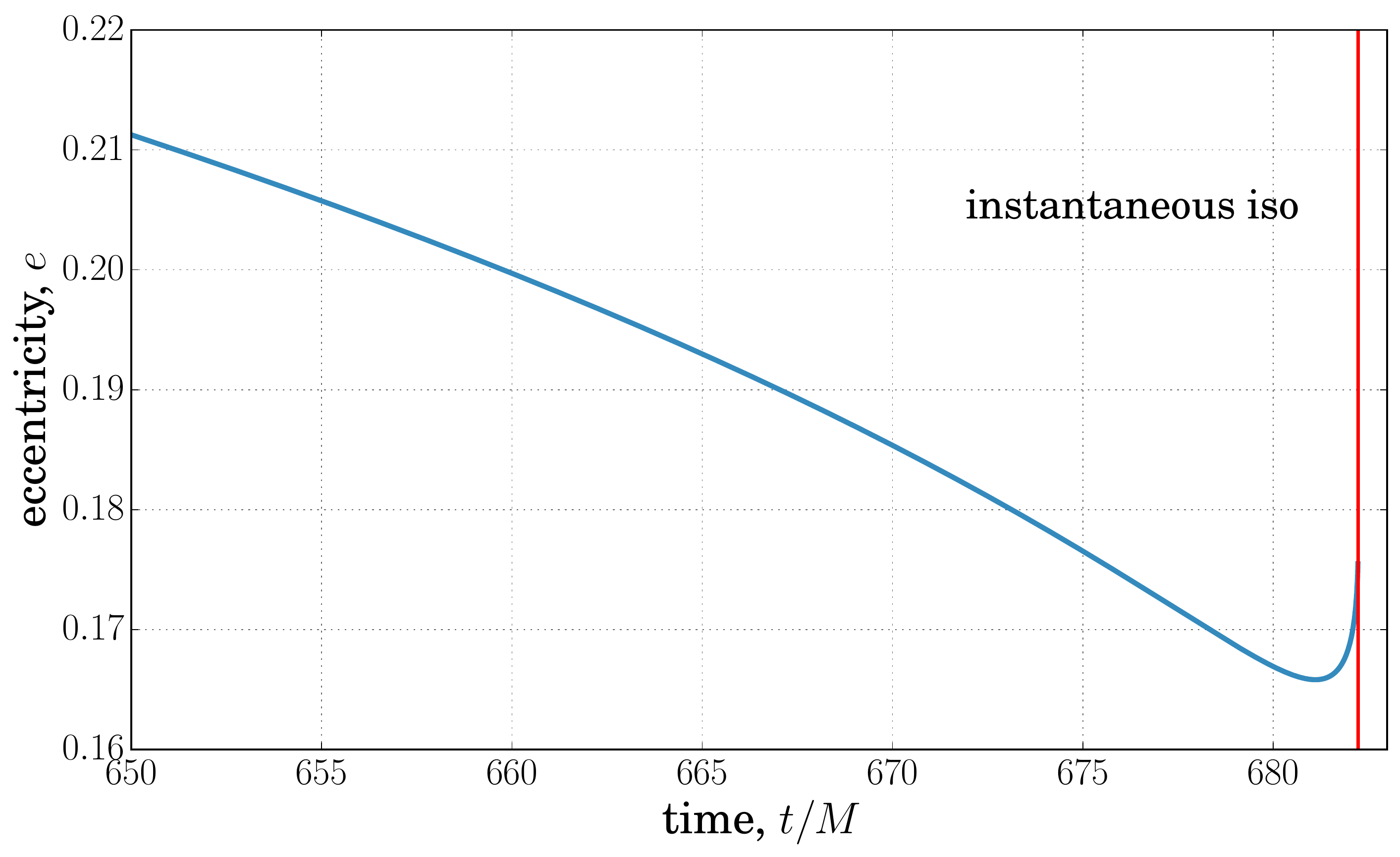}
\caption{\emph{Evolution of the eccentricity during an eccentric inspiral with mass ratio 4.} The vertical line marks the instant when trajectory reaches the instantaneous separatrix (iso) of the conservative dynamics. The initial eccentricity at $p=10$M was $e=0.4$, and it is the configuration corresponding to the potential illustrated in Fig.~\ref{fig:potentialsnapshot}}
\label{fig:eccincrease}
\end{figure}

A further similarity of the EOB to test-particle inspirals becomes apparent when considering the orbital evolution. For test-particle inspirals, once the trajectory approaches the vicinity of the instantaneous separatrix, the eccentricity increases instead of decreasing as in weak-field situations, see e.g. Ref.~\cite{Cutler:1994pb} for a discussion. Note that the eccentricity $e$ has a conventional meaning for the orbits that are close to Newtonian, and serves more like a convenient parametrization in the highly relativistic regime. For EOB evolutions within the approximations considered here we observe a similar effect. Figure~\ref{fig:potentialsnapshot} shows three snapshots of the instantaneous radial potential at different times during the evolution of a binary with mass ratio 4 and an eccentricity of $0.4$ at $p=10$M. The trajectory retains a sufficiently large eccentricity to transition to the plunge through the separatrix. The corresponding evolution of the eccentricity, shown in Fig.~\ref{fig:eccincrease}, is qualitatively similar to what would be expected for a test-particle inspiral: just before reaching the instantaneous separatrix, marked as the vertical line in the plot, the eccentricity starts to increase rather than decrease. 


We now consider the inspiral trajectory and $(2,2)$ mode of the waveform. Higher modes will become increasingly important as the eccentricity increases but we leave an analysis of the spectrum to future work when we include higher-order PN and the test-particle information in the EOB radiative sector. For the purposes of illustrating features of the EOB inspiral trajectories and waveforms, we use the explicit results for the fluxes and $h_{22}$, where the tail contributions were computed in the low-eccentricity and PN approximation. We further include a smooth connection to merger-RD signals to demonstrate that the model can produce full waveforms and as a basis for future refinements.

\begin{figure*}
\includegraphics[width=0.9\textwidth]{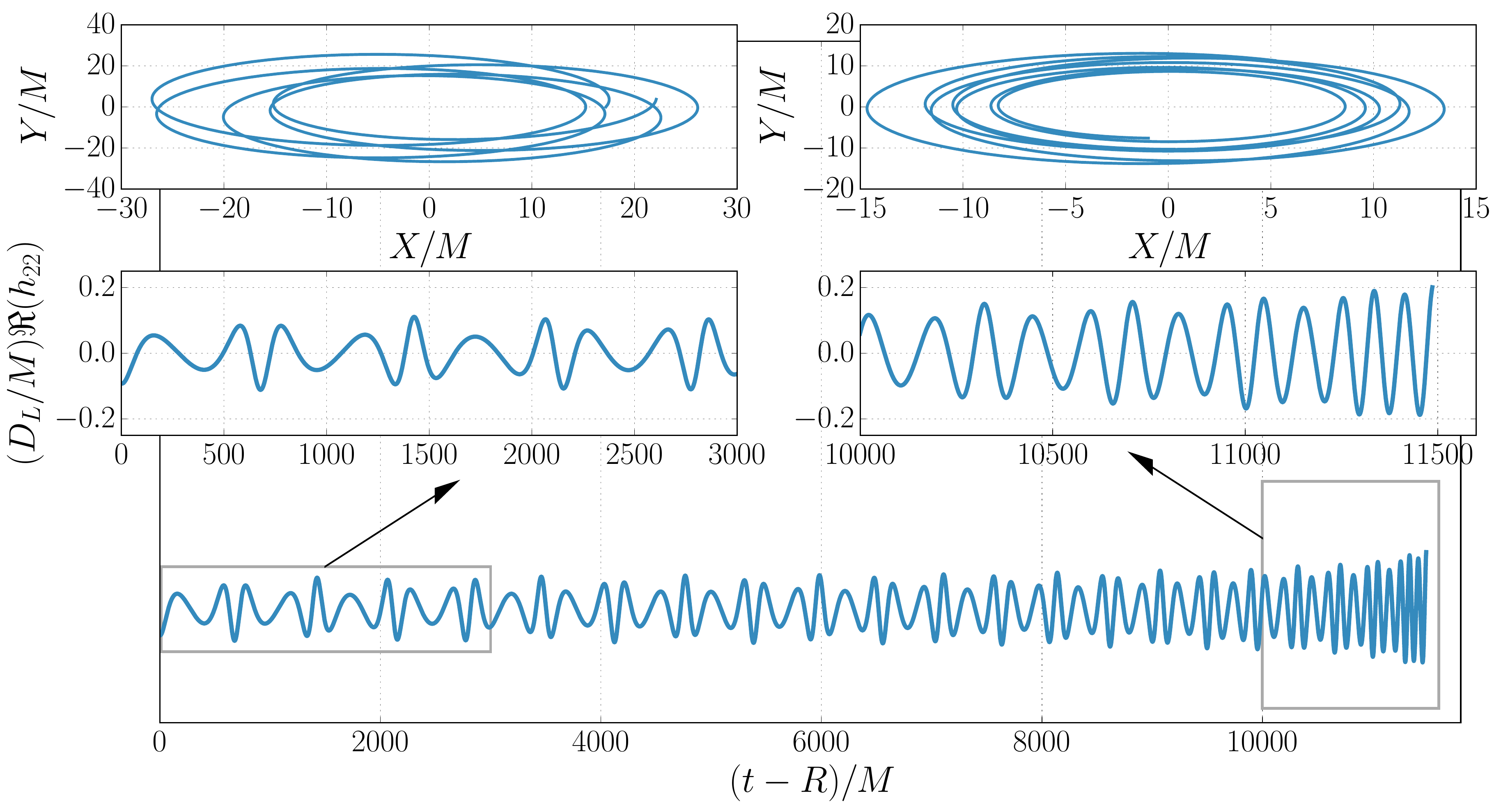}
\caption{\emph{Trajectory and $(2,2)$ mode of the waveform for an eccentric equal-mass inspiral in the adiabatic approximation.} The initial eccentricity was $0.3$ at $p=20$M but rapidly decreases during the evolution. }
\label{fig:inspiral}
\end{figure*}

Figure \ref{fig:inspiral} shows the waveform from a binary with an eccentricity of $0.3$ at $p=20M$. The blue curves are results from the rudimentary EOB model considered in this paper.
The insets show short traces of the trajectory at an early time in the evolution and a late time, when the system has already shed much of its eccentricity. This illustrates qualitatively the impact of eccentricity on the waveform and the orbital precession.

\begin{figure}
\includegraphics[width=\columnwidth]{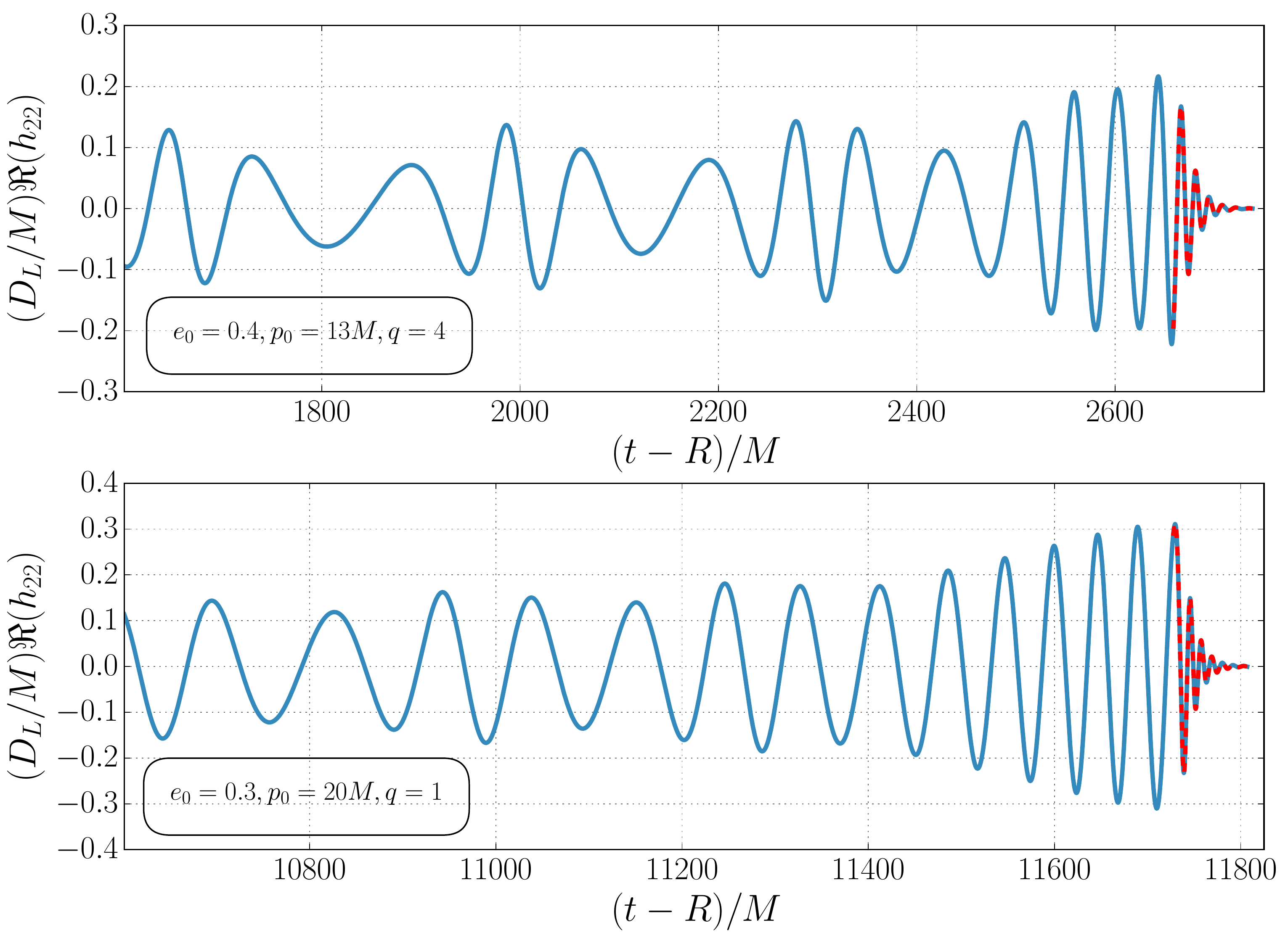}\caption{\emph{Examples of full Inspiral-Merger-Ringdown waveforms}. The setup shown in the upper  panel was chosen specifically to have a sufficiently large mass ratio and eccentricity that it passes through the separatrix. The lower  panel shows a configuration that has nearly circularized by the time of the merger. The merger-RD description is taken from the EOB model for circular inspirals.}  
\label{fig:IMRwaves}
\end{figure}

Figure~\ref{fig:IMRwaves} shows the completion of the inspiral signal with a circular merger-RD signal as described in Sec.~\ref{sec:MR}.  The upper panel  in Fig.~\ref{fig:IMRwaves} shows the results for a binary with mass ratio 4 that still has a non-negligible eccentricity at the end of the inspiral. In this case we also model the transition to the plunge and build again a complete waveform by attaching a merger-RD following procedure outlined 
in Sec~\ref{sec:MR}. The  lower  panel in Fig.~\ref{fig:IMRwaves} corresponds to a system that has already circularized during the inspiral, as is expected for most comparable-mass binaries visible to LIGO. Although the inspiral-plunge part computed here lacks  re-summed, higher order PN information  and the inputs from NR on the shape of the amplitude and frequency that are all part of the EOB model for quasi-circular binary coalescences, the plot demonstrates~\ref{fig:IMRwaves} that it is nevertheless possible to smoothly attach the merger-RD signal and obtain a full waveform.

\section{Conclusions and Outlook}
\label{sec:conclusion}

In this paper we have developed the foundations for an EOB model for the dynamics and gravitational waves from eccentric binary inspirals. Our approach employs an efficient parameterization of the dynamics that is adapted to the orbital geometry for relativistic binaries. Instead of working with the EOB coordinates and momenta the binary's degrees of freedom are divided into a set of phase variables and a set of quantities that are constant in the absence of radiation reaction and defined by the turning points of the radial motion. We derived the EOB dynamics in this parameterization and discussed the fundamental properties of the motion. Based on these insights we re-computed the fluxes of energy and angular momentum and the gravitational waveform from the expressions in terms of radiative multipole moments within the multipolar post-Minkowksi approximation. To calculate the instantaneous terms in the fluxes we started from explicit PN expressions in ADM coordinates and transformed them to EOB coordinates instead of using the multipole moments; for the Newtonian contributions we verified that these two methods lead to equivalent results in the appropriate limit.  For simplicity, we limited our explicit derivations of the gravitational radiation to $1.5$PN order, which already enabled us to discuss the general treatment of instantaneous and hereditary effects within our approach. We pointed out the greater transparency of the formulation used here  that is based directly on the frequencies of the motion  compared to QK parameterizations, and discussed resulting simplifications in the calculations. An important point to note is that the general procedure adopted here does not have any limitation to small eccentricity, even though many of the examples we considered were computed in the low-eccentricity approximation. We further addressed subtleties with the numerical implementation of our method and demonstrated the strategy to attach the merger-RD portion of the GW signal. Finally, we presented illustrative results for (i) quantities characterizing the dynamics, where no approximations were made, and (ii) for the gravitational waves that relied on several approximations, including the adiabatic limit, fluxes and waveforms computed for PN dynamics, to low PN order, and tail effects for low eccentricity, and with the merger-RD signals for circularized binaries. This provided a proof of principle that our approach is capable of describing all aspects of an eccentric binary inspiral and merger, and laid out the inputs and computational methods required for further refinements of the model.

In future work we will advance this model in several ways. It will be necessary to incorporate the knowledge of the fluxes and waveform to all known PN orders and to use the higher PN order non-geodesic terms in the Hamiltonian. These advances can be accomplished with the methods discussed in this paper but will require careful calculations. Future work is also needed on the possibility to use results from the gravitational self-force formalism. While this is straightforward in theory, the practical use requires an appropriate mapping that remains to be determined; for a description of the issues see e.g. the Appendix of Ref.~\cite{Steinhoff:2016rfi}.

For adiabatic waveforms, where the orbit-averaged radiation reaction forces are used for the evolution, we mentioned that existing PN results for the fluxes can readily be included in the model without repeating the calculations, by determining the mapping between eccentricity parameters from the PN expansion of the EOB energy and angular momentum. However, a main part of the remaining work is to compute the gravitational waveform modes and use them to construct the fluxes, to render the model self-consistent. The future work on the waveform modes will also require determining the appropriate EOB factorization of the PN results, and a re-summation of the hereditary terms based on the test-particle limit. As mentioned in the text, recent work~\cite{Forseth:2015oua} has examined analytical re-summations in the test-particle limit for the energy flux, however, this is not yet sufficient information to construct the factorized $h_{\ell m}$ modes in the EOB model. 

A key aspect of future work will be to test and improve the model by comparing to results from NR simulations. 
This will also serve to assess the performance of adiabatic waveforms. We plan to implement the completed model in the LIGO Algorithms Library for use in data analysis studies. Further effort will then be needed to incorporate spin effects in our approach. This work can make use of the existing EOB framework for modeling strong-field spin effects, and employ an extension of the methods developed for nonspinning systems in this paper, by identifying and computing the fundamental frequencies, associating a phase variable to each, and using these phases to perform Fourier decompositions of all the quantities needed to obtain the gravitational radiation. 

\acknowledgments
We thank Alessandra Buonanno and Maria Haney for a critical reading of the manuscript and for many useful comments. T. H. acknowledges support from the Radboud Excellence Initiative and partial travel support from ``NewCompStar", COST Action MP1304.

\appendix

\appendix

\section{EOB potentials}
\label{sec:EOBpotentials}

In Section~\ref{sec:implementation} we gave the A-potential in a general form in Eq.~\eqref{eq:Adecomp} that we reproduce here for convenience:
\be
A=\sum_{k=0}^{k_{\rm max}} a_k u^k+\left[b+g\,u^s \right] \log\left[f(u)\right], \label{eq:AdecompApp}
\ee
Below, we provide the coefficients and functions for the uncalibrated and the log-resummed, calibrated versions of the potential.

\subsection{Uncalibrated, Taylor-expanded potentials}
The A-potential is given by the expression from Eq.~\eqref{eq:AdecompApp} with $
k_{\rm max}=5$, $\, b=0$, $\,s=5$, and $ f(u)=u$. The coefficients are
\bes
\bea
a_0&=&1, \ \ \ \ a_1=-2, \ \ \ \ a_2=0 \ \ \ \ a_3=2\nu, \\
a_4&=& \nu \left(\frac{94}{3}-\frac{41\pi^2}{32}\right)\\
a_5 &=& \nu\left(\frac{128\gamma_E}{5}-\frac{4237}{60}+\frac{2275\pi^2}{512}+\frac{256\log(2)}{5}\right)\qquad\\
g&=& \frac{64}{5}\nu.
\eea
\ees
The functions $dA_j$ that appear in the regularized expression near the turning points in Eq.~\eqref{eq:Yapprox} are
\bea
dA_j^{\rm Taylor}&=& \sum_{k=0}^4 a_k \sum_{\ell=0}^{k-1} u^\ell u_j^{k-\ell-1}\nonumber\\
&+&\left[a_5+\frac{g}{2}\log(u u_j)\right]\sum_{k=0}^{4} u^k u_j^{4-k}\nonumber\\
&+&\frac{g}{2}(u^5+u_j^5)u_2^{1-j}u^{j-2}\sum_{k=0}^\infty\frac{(-1)^k}{(k+1)}\tilde\Delta_i^k, \qquad
\eea
where 
\be
\tilde \Delta_1=\frac{u_1-u}{u}, \ \ \ \ \ \ \ \ \tilde \Delta_2=\frac{u-u_2}{u_2}. \label{eq:deltatilde}
\ee

The $D$-potential is 
\bes\bea
D_{\rm Taylor}&=& 1+6\nu u^2+2 \nu u^3(26-3\nu)\nonumber\\
&&+\nu [d_4+d_{4l}\log(u)] u^4,
\eea
with
\bea
d_4&=& -\frac{533}{45}+\frac{1184\gamma_E}{15}-\frac{23761\pi^2}{1536}
-260\nu+\frac{123\pi^2\nu}{16}\nonumber\\
&&-\frac{6496\log(2)}{15}+\frac{2916\log(3)}{5}\\
 d_{4l}&=& \frac{592}{15}.  \label{eq:dcoeffs}
\eea
\ees
\subsection{Log-resummed potentials}
\label{sec:logpot}
The log-resummed, calibrated A-potential is given by Eq.~\eqref{eq:AdecompApp} with $k_{\rm max}=1$, $\, s=1$, and
\bea
&& a_0=\frac{1+\nu c_0}{(1- \nu K)^2}, \ \ \ \ \  a_1=-2 a_0 (1-\nu K),  \\
&& b=\frac{\nu}{(1- \nu K)^2}, \ \ \ \ \ g=-\frac{2\nu}{(1- \nu K)}.
\eea
The parameter $K$ is a calibration coefficient whose most recently updated value was determined in Eq.~(4.8) of Ref.~\cite{Bohe:2016gbl}. For this potential the function $f(u)$ is
\be
f(u)=1+\sum_{k=1}^5 c_k u^k+c_{5l}u^5 \log(u). 
\ee
The coefficients $c_k$ are written out explicitly in Appendix A of Ref.~\cite{Steinhoff:2016rfi}.
The functions $dA_j$ that appear in the regularized expression near the turning points in Eq.~\eqref{eq:Yapprox} are given by
\bea
dA_j(u) &=& a_1+\frac{g}{2}\log[f(u)\, f( u_j)]\nonumber\\
&+&\left[b+\frac{g}{2}(u+u_j)\right] Z_j \sum_{k=0}^\infty\frac{(-1)^k}{(k+1)}\Delta_j^k, \qquad
\eea
where $\Delta_1=(f_1-f)/f$, $\, \Delta_2=(f-f_2)/f_2$, and
\bea
Z_j&=& f^{j-2}f_j^{1-j}\bigg\{\sum_{k=1}^4\sum_{\ell=0}^{k-1} c_k u^\ell u_j^{k-\ell-1} \\
&& \qquad+\frac{c_{5l}}{2}\sum_{\ell=0}^{4}u^\ell u_j^{3-\ell} \log(u\, u_j)\nonumber\\
&& \qquad+ \frac{c_{5l}}{2} (u+u_j)u_2^{1-j}u^{j-2}\sum_{k=0}^\infty\frac{(-1)^k}{(k+1)}\tilde\Delta_j^k\bigg\}. \nonumber
\eea 
The $D$-potential is
\be
D=1+\log\left[D_{\rm Taylor}\right].\ee

\subsection{Regularizing the radiation reaction terms in the equations of motion}
\label{sec:rrterms}
In this Appendix we discuss the numerical treatment of the non-geodesic terms appearing in~\eqref{E:rad_mot}. These are relevant when going beyond the adiabatic approximation discussed in the body of the paper. As in the case of the conservative dynamics, the factor $(\partial r/\partial\xi)^{-1}$ introduces an apparent divergence near the turning points. It is therefore necessary to compute an expression for the term $\dot{C}_i \partial r/\partial C_i$ in Eq.~\eqref{E:rad_mot} that manifestly cancels this divergence. To obtain an expression for $\dot{C}_i \partial r/\partial C_i$ we differentiate the radial potential, defined in Eq.~\ref{eq:Vrdef}, in two ways: first considering $d/d C_i$ and then $d/dt$. These operations lead to the following relations respectively
\bes
\label{eq:usefulderivs}
\bea
2\dot{r}\frac{d \dot{r}}{d C_{i}} \dot{C}_i &=& \frac{\di V_r}{\di C_i}\dot{C}_i + \frac{\di V_r}{\di r} \frac{\di r}{\di C_i} \dot{C}_i,\\
2 \dot{r}\ddot{r} &=& \frac{\di V_r}{\di r}\dot{r} + \frac{\di V_r}{\di C_i} \dot{C}_i, 
\eea
where we have multiplied the first expression by $\dot C_i$. We also note that the radial acceleration due to radiation reaction is
\be
\label{eq:radialacc}
2\dot{r}a^r = 2\dot{r} \left[ \ddot{r} - \frac1{2}\frac{\di V_r}{\di r} \right]=\frac{\di V_r}{\di C_i} \dot{C}_i.
\ee
\ees
We solve Eqs.~\eqref{eq:usefulderivs} for $(\partial r/\partial C_i) \dot{C}_i$ and obtain
\be
\dot{\xi} ={\cal P}- \frac{2\dot{r}}{\di r/\di \xi} \frac1{\di V_r/\di r}
\left[ 
\frac{d\dot{r}}{dC_i} \dot{C}_i - a^r
\right].
\label{E:rad_mot_2}
\ee
This expression~\eqref{E:rad_mot_2} is in the desired form that is manifestly finite at the turning points. 
Specifically, each of the factor is finite for the following reasons. The first factor is directly related to the quantity $\dot{r}/\di r/ \di \xi ={\cal P}$ that we computed near the turning points in the previous subsection. For the second factor, $(\di V_r/\di r)^{-1}$, we note that from Eq.~\eqref{eq:Vrdef} that the radial derivative of $V_r$ evaluated at the turning points is 
\be
\frac{\di V_r}{\di r}\bigg\rvert_{r_{1,2}} = X^2(r_{1,2}) \frac{\di}{\di r}\left[\frac{2A^{-1}}{\beta D}\left(\sqrt{1+\beta Y}-1\right)\right]_{r_{1,2}}
\ee
which is non-zero.

To show that the expression inside the square brackets of \eqref{E:rad_mot_2} is regular we consider its expansion near the turning points. Since $\dot r=X \hat P_r$, we see immediately from Eqs.~\eqref{eq:prnearturning} and~\eqref{eq:Yapprox} that it is of the form $\dot r\propto \sin\xi $, where the proportionality factor depends on $(r,C_i)$ but is not needed explicitly here. Differentiation then leads to an expression of the form $\partial \dot{r}/\partial C_i = \sin{\xi}\, \partial/{\partial C_i}\left(\ldots\right)$,
which is regular at turning points. A similar argument applies for the computation of the radial acceleration from Eq.~\eqref{eq:radialacc}, since $\dot E$ and $\dot P_\phi$ are related to the fluxes and remain divergence-free at the turning points. 

However, the numerical problem with the expression~\eqref{E:rad_mot_2} is that $\di V_r/\di r$ goes through zeros in between the turning points, as can be seen from the fact that $V(r_{1,2}) = 0$ at both turning points. 
Hence, the idea for the practical implementation is to switch between two representations: (i) the original expression 
\eqref{E:rad_mot} for the portion of the dynamics away from the turning points, and \eqref{E:rad_mot_2} for the dynamics near the turning points. This technique was also used in \cite{Gair:2010iv}. Specifically, the prescription is
\be
\label{eq:xidotfinal}
\dot{\xi} = {\cal P} - 
\left\{
\begin{array}{cc}
\left(\frac{\partial r }{\partial \xi} \right)^{-1}\frac{\di r}{\di C_i} \dot{C_i} & r_2\ll r\ll r_1 \\
2{\cal P} \left(\frac{\di V_r }{\di r}\right)^{-1} \left[ \frac{d \dot{r}}{dC_i}\dot{C}_i - a^r \right] &  r \, {\rm near} \, r_{1,2}
\end{array}
\right.
\ee

\bibliography{inspire}
\end{document}